\UseRawInputEncoding
\RequirePackage{fix-cm}
\documentclass[smallextended]{svjour3}       % onecolumn (second format)
\smartqed  % flush right qed marks, e.g. at end of proof
\usepackage{graphicx}
\usepackage{amsmath,amssymb,amsfonts}
\usepackage{multirow}
\usepackage{lscape}
\usepackage{longtable}
\usepackage{supertabular}
\usepackage{subfig}
\usepackage{natbib}
\usepackage{enumitem}
\usepackage{algorithmic}
\usepackage{graphicx}
\usepackage{textcomp}
\usepackage{xcolor}
\usepackage{pgfplots}
\usepackage{caption}
\usepackage{tikz}
\usepackage{pgfplotstable}
\usepackage{comment}
\usepackage{booktabs}
\usepackage{tcolorbox}
\usepackage{url}
\usepackage{balance}
\usepackage{color, colortbl}
\usepackage{tabularx}
\usepackage{adjustbox}
\usepackage{diagbox}
\usepackage{wrapfig}
\usepackage[colorinlistoftodos]{todonotes}

\usepgfplotslibrary{statistics}

\definecolor{gray}{gray}{0.85}
\definecolor{cyan}{rgb}{0.88,1,1}
\definecolor{ao}{HTML}{008000}
\definecolor{lava}{HTML}{CF1020}

\newlength{\maxlen}

\newcommand\fix[1]{{\textcolor{black}{#1}}}

\newcommand{\RqOne}{\textbf{\emph{RQ$_1$: \fix{How does participation differ according to membership status?}}}}
\newcommand{\RqTwo}{\textbf{\emph{RQ$_2$: \fix{How do communication content vary for different membership statuses?}}}}
\newcommand{\RqThree}{\textbf{\emph{RQ$_3$: \fix{What are the differences perceived in the sentiment of interaction based on membership status?}}}}

\begin{document}

\title{How are Project-Specific Forums Utilized? A Study of Participation, Content, and Sentiment in the Eclipse Ecosystem}


\author{Yusuf Sulistyo Nugroho         \and
        Syful Islam \and
        Keitaro Nakasai \and
        Ifraz Rehman    \and
        Hideaki Hata	\and
        Raula Gaikovina Kula \and
        Meiyappan Nagappan  \and
        Kenichi Matsumoto
}


\institute{Yusuf Sulistyo Nugroho \at
            Universitas Muhammadiyah Surakarta, Indonesia   \\
            \email{yusuf.nugroho@ums.ac.id}\\
            \and
            Syful Islam,  
            Ifraz Rehman, Raula Gaikovina Kula, Kenichi Matsumoto \at
            Nara Institute of Science and Technology, Japan \\
            \email{\{islam.syful.il4, rehman.ifraz.qy4, raula-k, matumoto\}@is.naist.jp} \\
            \and
            Keitaro Nakasai \at
            National Institute of Technology, Kagoshima College, Japan\\
            \email{nakasai@kagoshima-ct.ac.jp} \\
            \and
            Hideaki Hata \at
            Shinshu University, Japan \\
            \email{hata@shinshu-u.ac.jp} \\
           \and
            Meiyappan Nagappan
            \at University of Waterloo, Canada    \\
            \email{mei.nagappan@uwaterloo.ca} \\
}

\date{Received: date / Accepted: date}

\maketitle

\begin{abstract}
\fix{
Although many software development projects have moved their developer discussion forums to generic platforms such as Stack Overflow, Eclipse has been steadfast in hosting their self-supported community forums.
While recent studies show forums share similarities to generic communication channels, it is unknown how project-specific forums are utilized.
In this paper, we analyze 832,058 forum threads and their linkages to four systems with 2,170 connected contributors to understand the participation, content and sentiment.
Results show that Seniors are the most active participants to respond bug and non-bug-related threads in the forums (i.e., 66.1\% and 45.5\%),
and sentiment among developers are inconsistent while knowledge sharing within Eclipse.
We recommend the users to identify appropriate topics and ask in a positive procedural way when joining forums.
For developers, preparing project-specific forums could be an option to bridge the communication between members.
Irrespective of the popularity of Stack Overflow, we argue the benefits of using project-specific forum initiatives, such as GitHub Discussions, are needed to cultivate a community and its ecosystem.
}
\end{abstract}

\section{Introduction}
\label{intro}
As one of the discussion group channels, forums support mass communication and coordination among distributed software development contributors~\citep{Storey:2017:SCC:3057931.3057955}.
Forums are considered to be an improvement over mailing lists in that it provides browse and search functions, which are especially helpful for repetitive questions and answers~\citep{Squire:2015:WMS:2819009.2819042}.

Recent studies have been conducted that compare forums to communication channels that seem to have similarities to a forum.
For instance, there have been studies on mailing lists~\citep{Guzzi:2013:COS:2487085.2487139,Zagalsky:2018:RCC:3211160.3211168},
question and answer sites (Stack Overflow)~\citep{Ye:2017:SDK:3042021.3042050,Wang:2018:UFF:3231288.3231332,Zou:2017:TCN:3044551.3068580,Calefato:2018:ATH:3163583.3163673},
Microblogs~\citep{Guzman:7765515,Mezouar:2018:TUB:3231288.3231333}, and
News aggregators~\citep{Aniche:2018:MNA:3180155.3180180}.
Kahani et al. analyzed discussion topics using a topic modeling technique~\citep{Kahani:2016:PEM:2976767.2976773}.
Squire studied the transition from self-supported forums to Stack Overflow~\citep{Squire:2015:WMS:2819009.2819042}.
It has been reported that generic  question and answer platforms such as Stack Overflow have taken over the roles of forums.
For instance, many software development projects had closed their self-supported forums and moved to Stack Overflow~\citep{Squire:2015:WMS:2819009.2819042}.
Furthermore, gamification strategies such as awarding of badges or a voting system of Stack Overflow are considered to be incentives designed to participate and improve answer quality~\citep{Squire:2015:WMS:2819009.2819042}.

\fix{From the perspective of forums, the novelty of this work is a comprehensive investigation into how a large forum in the Eclipse ecosystem is utilized.
Due to the large amount of information available, we are now able to cover different facets of communication between its members. 
As a long-living free and libre and open-source software (FLOSS) project, the Eclipse project still maintains an established and active forum, that has also been a targeted data source for three MSR Mining Challenges.\footnote{MSRMiningChallenge2007: \url{http://2007.msrconf.org/challenge/}.}$^{,}$\footnote{MSRMiningChallenge2008: \url{http://2008.msrconf.org/challenge/}.}$^{,}$\footnote{MSRMiningChallenge2011: \url{http://2011.msrconf.org/msr-challenge.html}.}
The forum handles its massive ecosystem, centralizing all contributor activities within the Eclipse ecosystem, such as contributed projects, reviews in Gerrit, and topics in forums (i.e., Profile, Forums, Gerrit, Bugzilla, Projects).\footnote{Antoine Thomas, The Eclipse User Profile,  \url{http://blog.ttoine.net/en/2016/12/01/the-eclipse-user-profile/}, December 1, 2016.}
Furthermore, Eclipse forums support donation badges, which does have a practical impact and signalling on decreasing response time of bug reports~\citep{8501934}. 
In addition, unlike the other online question and answer platforms such as Stack Overflow, Eclipse manages their users, assigning different contributions based on status (i.e., Junior, Member, and Senior).}

In this paper, we conduct an investigation the Eclipse forum \fix{in terms of participation, content and sentiment}.
We first perform an analysis to classify the membership status of the users.
Then we set out to empirically investigate forum threads and 2,170 connected contributions to other four systems, (i.e., around 416 thousand profiles, 120 thousand code review submissions, 532 thousand bug reports with 2,883 committers from multiple projects) within the Eclipse ecosystem. To guide our empirical analyses, we formulated the following research questions:
\begin{itemize}
    \item \RqOne
    
    \textit{Motivation:} First, this study is conducted to answer the extent to which users in the ecosystem participate in (1) using forums and (2) using various systems. 
    Second, we would like to analyze the extent to which different types of forum members in the ecosystem take to respond a particular type of post.    
    \\
    \textit{Results:} Senior Members are the most active forum users to participate in responding to the bug-related threads (66.1\%) and the non-bug-related threads (45.5\%), with active contributors in other systems are also active in forums, making it a source of expert knowledge for all systems. 

    \item \RqTwo
    
    \textit{Motivation:} The key motivation for \textit{RQ$_2$} is to understand the message contents posted by both the organization and users, the most frequent content of posts in the forums, and how users reference the information sources to support their answers. 
    In detail, we also investigate different contents that users share in Eclipse forums.  
    \\
    \textit{Results:} Eclipse forums are dominated by question and answer threads, especially discrepancies, amounting to 28.5\%, most of which were posted by Junior Members, with replies and welcome messages being the two most frequent posts by the organization.

    \item \RqThree 
    
    \textit{Motivation:} The motivation of {RQ$_3$} is to get deep insight on developers sentiment while sharing knowledge. 
    Throughout this {RQ}, we aim to test the degree of consistency while interacting with each other.
    \\
    \textit{Results:} The sentiment among developers is inconsistent while knowledge sharing in the Eclipse forums.
    
\end{itemize}

\fix{
Our results provides arguments for the need to support project-specific forums, especially for large ecosystems of projects.
In summary, the main contributions of our work are as follows (i) a comprehensive study of threads in Eclipse forum, covers 832,058 threads with links to profiles, code review, bugs and project systems, (ii) an analysis of Eclipse forum membership using topological analysis, (iii) a manually labelled taxonomy of forum content communicated, (iv) a manual analysis on social interactions within an ecosystem and (v) a set of recommendations for both researchers, and practitioners on the use of forums.
}

The rest of this paper is structured as follows.
Section~\ref{sec:datacollection} describes our procedure to collect data through five different systems in the Eclipse ecosystem, membership classification, and an online appendix.
Section~\ref{sec:method} presents our approach to empirically analyze the users' participation, content of forums, and sentiment of interactions.
The results of the study are presented in Section~\ref{ssec:mainresults}.
Section~\ref{sec:recommendation}, \ref{sec:threatstovalidity} and \ref{sec:relatedwork} describe our recommendation, threats to validity and provide related works.
Finally, we conclude this paper in Section~\ref{sec:conclusion}.

\section{Data Collection}
\label{sec:datacollection}
In this section, we describe the procedure of our data collection, and present our replication package. 
Our goal is to investigate the nature of participation, content, and sentiment of interaction based on the membership in the Eclipse forums.
Thus, in this study, we also present how we distinguished the statuses of members. 

\subsection{Eclipse Community System Collection}
\label{ssec:cdc}

\begin{figure*}
    \centerline{\includegraphics[width=1\linewidth]{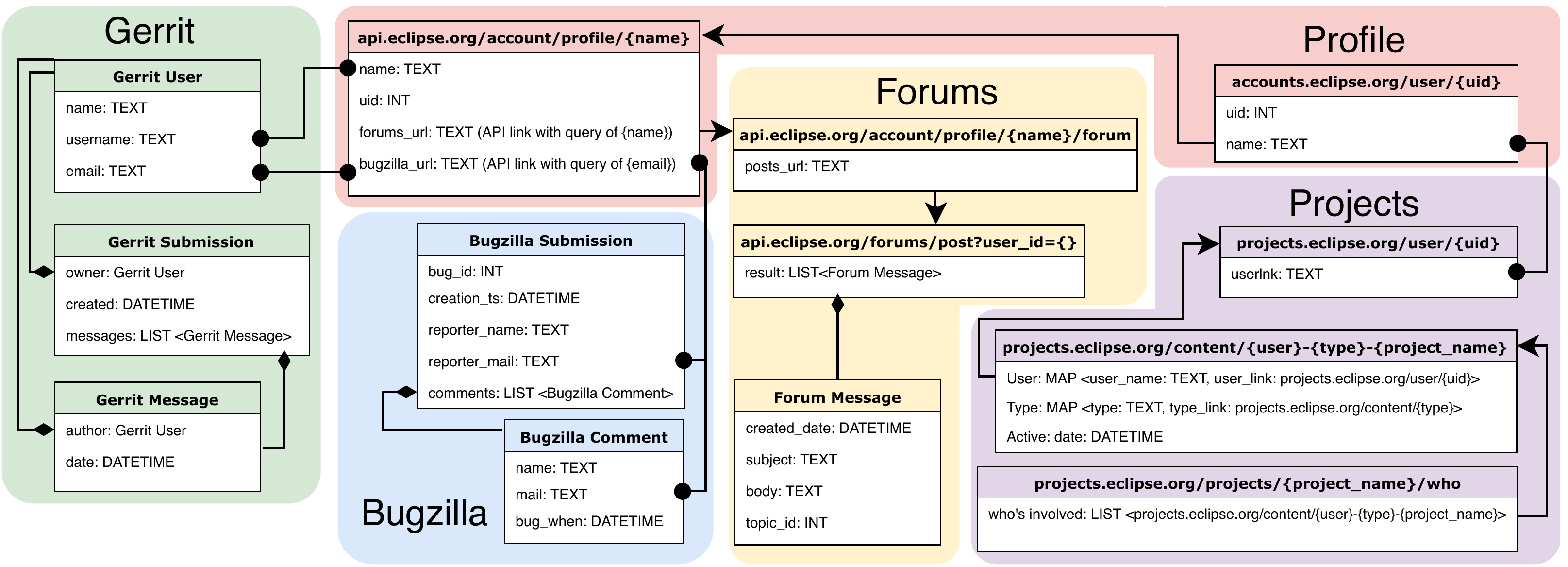}}
    \caption{Connected data from Gerrit, Bugzilla, Forums, and Projects via REST API in the Eclipse ecosystem.}
    \label{fig:datalink}
\end{figure*}

Figure~\ref{fig:datalink} illustrates the schema that connects the five software development datasets within the Eclipse ecosystem.
This connected data enables the integration of contributors' various activities.
Since the Eclipse REST API requests are limited to 1,000 an hour, we collect data not only from API but also from individual systems, 
that is, Profile, Forums, Gerrit, Bugzilla, and Eclipse Projects.

\begin{itemize}
    \item 
    \textbf{Profile Dataset}. 
    Each contributor in the Eclipse ecosystem is required to register for a profile.\footnote{\url{https://accounts.eclipse.org/user/register}}
    The profile contains basic contributor information and can be used to track contributor activities.

    \item 
    \textbf{Bugzilla Dataset}.
    The Bugzilla dataset contain the tracking of bug reports and their fixes within the Eclipse ecosystem.\footnote{\url{https://bugs.eclipse.org/bugs/}}

    \item 
    \textbf{Projects Dataset}. 
    To track social roles (committer, review, mentor, etc.)  of contributors, we include the project dataset.\footnote{A detailed list of Eclipse projects is available from \url{https://projects.eclipse.org/}.}

    \item 
    \textbf{Gerrit Dataset}. 
    Gerrit is a review tool that facilitates collaboration between committers and contributors \fix{within} Eclipse.
    
    \item 
    \textbf{Forum Threads}. 
    Eclipse community forums, a user-to-user interaction site for Eclipse users, have a hierarchical structure.
    It contains a number of sub-forum categories which may have several topics.
    Within a forum's problem-related topic, each new initial \fix{post} by a user can have a response from other users in the community.
\end{itemize}

The participation of contributors within the Eclipse ecosystem can be analyzed from the connected data of five datasets, as described in Figure~\ref{fig:datalink}.
As presented in Table~\ref{tab:fivedataextraction}, the data used in this analysis are extracted through several steps.

\begin{table}
    \centering
    \caption{Connected data extraction from five data sources within the Eclipse ecosystem}
    \label{tab:fivedataextraction}
    \begin{tabular}{llrl}
        \toprule
        \textbf{step} & \textbf{dataset} & \textbf{quantity} & \\
        \midrule
        step 1: & Profile & 416,126 & profiles  \\
         & Bugzilla & 531,752 & bug reports    \\
         & Projects & 2,883 & committers   \\
         & Gerrit & 120,165 & submissions \\
         & Forums & 1,097,174 & threads   \\
        step 2: & Connected Gerrit contributors & 2,170 & contributors \\
         & Connected forum messages & 467 & messages   \\
         
        \bottomrule
    \end{tabular}
\end{table}

\textbf{Step 1:} In this step, we extracted the dataset from five different sources, that are, Profile, Bugzilla, Projects, Gerrit, and Forums.
For Profile dataset extraction, 416,150 \textit{uid}s and \textit{name}s were collected on September 14, 2018 from the profile system.\footnote{\url{https://accounts.eclipse.org/user/}}
Using the Profile API, 416,126 Profile data were obtained (24 users were not found) from November 5 to 13, 2018.
To connect the activities in Bugzilla with the Profile dataset, we used the \textit{email} addresses.
We extracted 531,752 bug reports from October 10, 2001 to January 13, 2019.
\fix{From the Projects data source, we then collected a total of 2,942 committers from the obtained 438 projects (on January 16, 2019).}
Among them, the information of 59 committers could not be accessed because of the deletion of accounts.
Hence, the remaining 2,883 committers can be connected with the Profile dataset through the \textit{uid}.
We collected 120,165 Gerrit submissions from October 1, 2009 to October 31, 2018.
We link users with the Profile dataset, using the \textit{name} or \textit{email} to match.
To download the forum threads, we extract the data from its API. 
We downloaded 1,097,174 threads (topics) available in the Eclipse community forums until January 9, 2019.\footnote{\url{https://www.eclipse.org/forums/index.php/t/}}

\textbf{Step 2:} The connected forum data are collected in this step.
On November 23, 2018, we extracted forum message data from Gerrit contributors, via the Eclipse REST API. 
As part of the data collection process, we applied a pre-processing to detect and remove duplicated accounts.
We report 39 pairs of Gerrit and Eclipse forum accounts that have same \textit{name}s but different \textit{username}s. 
Through manual examination, we verified identities within those pairs; 27 pairs were found to be identical and merged.
After the duplicate removal and linking to Profile data, we obtained 2,170 Gerrit contributors and extracted 467 forum messages that connect to these Gerrit contributors. 

The collected 2,170 contributors from this extraction are then used to investigate the participation of contributors within the Eclipse ecosystem using a 
topological data analysis technique (i.e., TDA).
TDA is a method to extract valuable information from high-dimensional data that is insensitive to metric, noisy, and incomplete without formulating an initial hypothesis~\citep{Lum13}.
To build the topology, we construct the metrics of each contributor from the four data sources, namely, Bugzilla, Gerrit, Forums, and Projects (see the details in Section~\ref{sssec:approachPS1}).
The metrics describe the connected activities of the contributors within the Eclipse ecosystem.

\label{ssec:ftc}

\begin{figure}
    \centering
    \includegraphics[width=.87\linewidth]{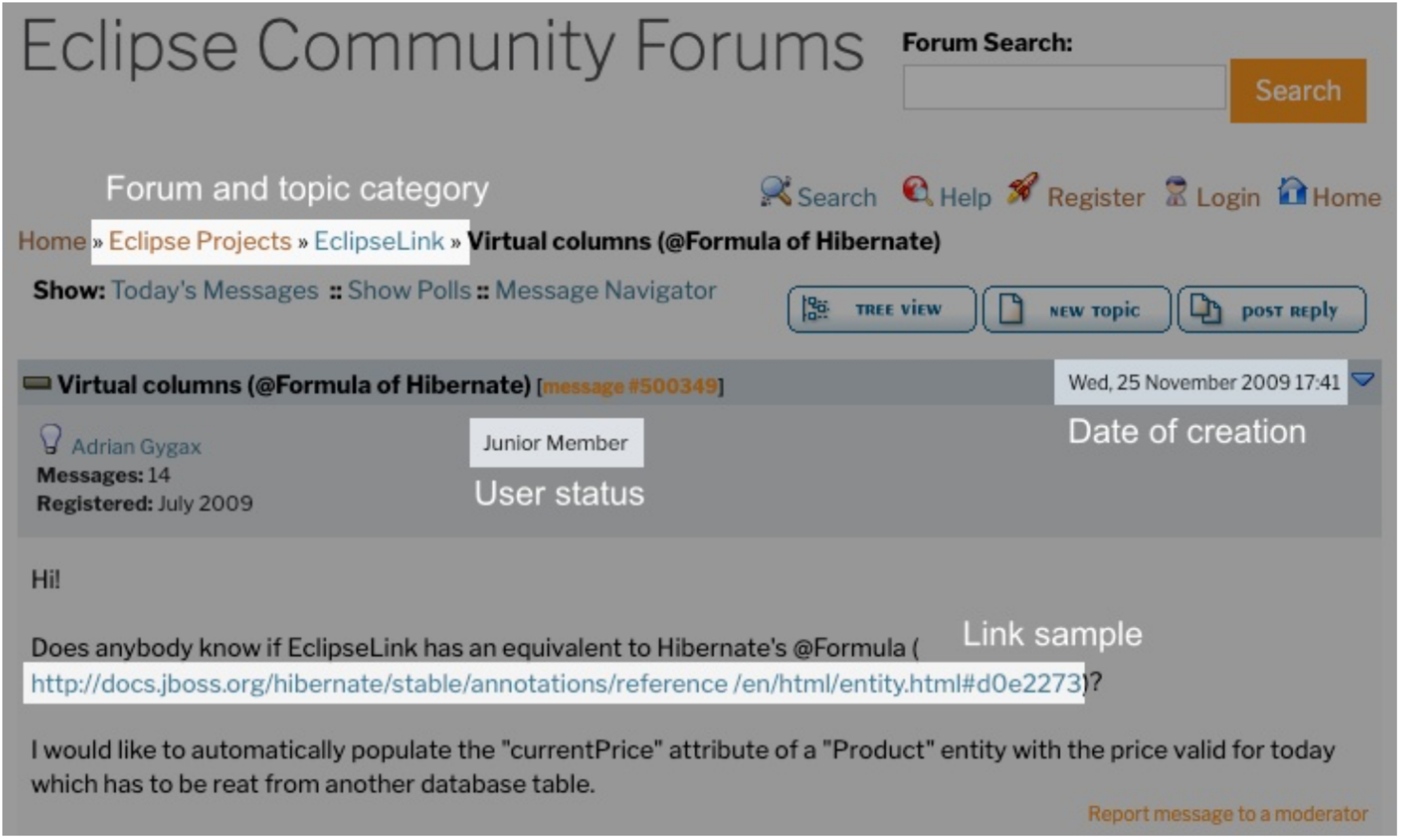}
    \caption{Example of a forum thread}
    \label{fig:threadsample}
\end{figure}

Figure \ref{fig:threadsample} depicts different elements of a thread in the Eclipse forums, including (i) the topic category and subcategory, (ii) the member status, (iii) date of creation, and (iv) the link that was posted in the thread.

To improve the quality of our dataset, as shown in Table \ref{tab:noofthreadeachstep}, the forum threads underwent three stages of filtering.

\begin{table}
    \centering
    \caption{Outputs of the pre-processing of the Forum Dataset}
    \label{tab:noofthreadeachstep}
    \begin{tabular}{llr}
        \toprule
        \textbf{step} & & \textbf{\# threads}\\
        \midrule
        step 1: & raw extraction & 1,097,174  \\
        step 2: & remove duplication & 832,058 \\
        step 3: & separation: &  \\
         & (i) threads by webmaster & 542,997 \\
         & (ii) threads by non-webmaster users & 289,061 \\
        \bottomrule
    \end{tabular}
\end{table} 

\begin{enumerate}
    \item \textbf{Step 1}. Raw extraction. In this step, we use all 1,097,174 collected threads yielded from \textit{Step 1} in Table~\ref{tab:fivedataextraction}.
    \item \textbf{Step 2}. Remove duplication. In the Eclipse community forums, we found that some threads had been duplicated by the system several times. 
    To avoid redundancies, we removed such threads based on the identity number of \fix{initial posts}. 
    We were able to reduce the number of threads to 832,058.
    \item \textbf{Step 3}. Separation. To investigate how the organization communicates with the members and analyze the characteristics of forum usages, contents, etc., we separate those threads posted by the webmasters and the non-webmaster users that include the threads from users that contribute in the other systems.
    Webmasters deal with the servers and software that runs the eclipse.org site.~\footnote{\url{https://wiki.eclipse.org/WebMaster}}
    Webmasters have different roles in the forum than other users, so they need to be classified. 
    Also, by analyzing the content of webmasters’ posts, it is possible to understand how the organization is engaged in the forum.
    From the total number of threads resulted in \textit{Step 2}, we were able to separate (i) 542,997 threads posted by the webmasters and (ii) 289,061 threads posted by non-webmaster users.
\end{enumerate}

\subsection{Membership classification}
\label{ssec:memberdefine}
This section describes our techniques to classify the membership of users for each posted message.
Unlike the other question and answer online forums such as Stack Overflow, in the Eclipse community forum, all registered users are assigned into
three statuses of membership, that are, (1) Junior, (2) Member, and (3) Senior.
These user statuses are included in our collected data resulted from \textit{Step 3} in Table~\ref{tab:noofthreadeachstep} which can be seen in every post of a user, as shown in Figure~\ref{fig:threadsample}.
The status of each user may change from the lowest level (i.e. Junior) into the highest one (i.e. Senior) depending on the contributions of the user in the community.
However, in the forum, we could not differentiate which posts were posted by users when they were a Junior, Member or Senior.
This is because once the status of a user has changed, it will replace the old status in all posts of a user with the latest status, including their first posts.
Furthermore, we also did not find any information about the time when the status of a user changed.

\begin{figure}
    \centering
    \begin{tikzpicture}
        \begin{axis}
        [
        ytick={1,2,3},
        yticklabels={Junior, Member, Senior},
        width=\textwidth,
        height=6.5cm,
        xmax=610,
        xmin=-10,
        xlabel={\# posts},
        nodes near coords align=right,
        every node near coord/.style={color=black},
        ]
        \addplot+[mark=x,
        nodes near coords=29,
        boxplot prepared={
          median=4,
          upper quartile=8,
          lower quartile=2,
          upper whisker=15,
          lower whisker=1
        },
        ] coordinates {(0,29)};
        \addplot+[mark=x,
        nodes near coords=106,
        boxplot prepared={
          median=40,
          upper quartile=56,
          lower quartile=32,
          upper whisker=92,
          lower whisker=30
        },
        ] coordinates {(0,106)};
        \addplot+[mark=x,
        boxplot prepared={
          median=163,
          upper quartile=296,
          lower quartile=116,
          upper whisker=562,
          lower whisker=107
        },
        ] coordinates {(0,570)(0,580)(0,590)(0,600)};
        \end{axis}
    \end{tikzpicture}
    \caption{Frequency of posts per user status. The maximum number of posts for each type of users is used to define the threshold of post-based membership. The threshold for Juniors and Members are 29 and 106, respectively. Although Seniors have posted more than 28 thousands posts, we limit up to 600 in the figure.}
    \label{fig:boxplotmembership}
\end{figure}
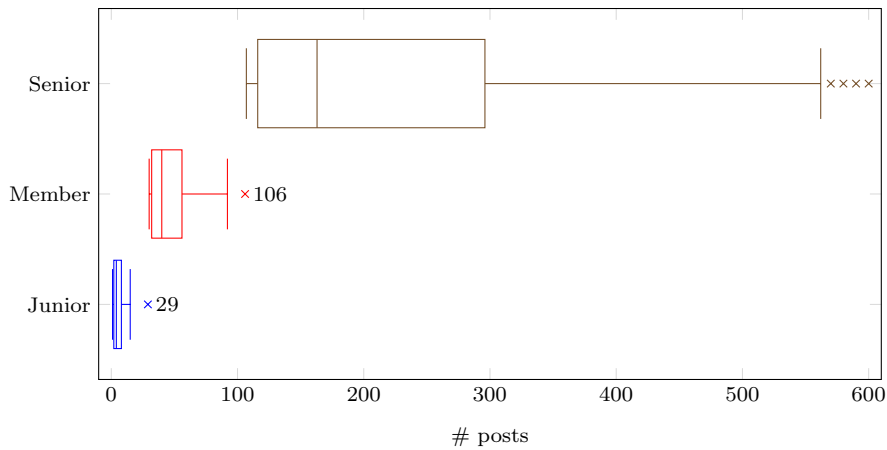

To define the member status of each registered user, we attempted to calculate the total number of posts of every user.
From this amount of \fix{posts}, we summarized the quantity of posts per author based on the user identity number.
The total number of posts per author varies, from less than ten to more than one thousand posts.
In this step, we found the maximum number of posts of each user if we consider the latest status of users as collected in the dataset, as shown in Figure~\ref{fig:boxplotmembership}.
The maximum number of \fix{posts} by Juniors and Members are 29 and 106 respectively, while the maximum number of posts by Seniors is 28,476. 
Based on this finding, we used these maximum numbers as the thresholds to differentiate the user status for each \fix{post} based on the sequence number of a post.
The sequence number of a post depends on its creation date in order.
The earliest post is assigned as the first post, then followed by the other posts ordered by date of creation.

\textbf{Users' classification.}
As shown in Table~\ref{tab:numberofrole}, we are able to assign the status of the users in all \fix{posts} based on the sequence number of the posts and able to distinguish the \fix{posts} that were posted by the users when they were Junior, Member and Senior.
The \fix{posts} that are assigned from 1 to 29 are considered as posts that were posted by Juniors, while posts from number 30 to 106 were posted by Members, and Seniors posted the \fix{posts} from 107 up to the remaining posts.

\begin{table}
    \centering
    \caption{Frequency of posted threads based on the classification results}
    \label{tab:numberofrole}
    \begin{tabular}{lrr}
        \toprule
        \textbf{status} & \textbf{\# threads} & \textbf{(\%)}    \\
        \midrule
        Junior Member & 144,440 & (50\%)  \\
        Member & 31,719 & (11\%)  \\
        Senior Member & 26,443 & (9\%) \\
        Eclipse User & 86,459 & (30\%)    \\
        \midrule
        \textbf{sum} & \textbf{289,061} & \textbf{(100\%)}  \\
        \bottomrule
    \end{tabular}
\end{table}

In the Eclipse forums, we also found that there are a lot of \fix{posts} posted by authors with the username ``Eclipse User'', accounting for 30\% of the forum users, as described in Table~\ref{tab:numberofrole}.
In the threads, this username is automatically assigned by the system and could be from anyone.
It does not show the status explicitly whether Junior, Member, or Senior.
Furthermore, it does not have a user identity number and a link that connects to the user's profile. 
Since we only focus to identify the three statuses of the user in the analyses, the messages that were posted by ``Eclipse User'' were excluded.
After the exclusion process, we obtained 202,602 threads that were only posted by Juniors, Members and Seniors.

\begin{figure*}
    \centering
    \subfloat[\# Junior Members vs. \# posts]{\includegraphics[width=.5\textwidth]{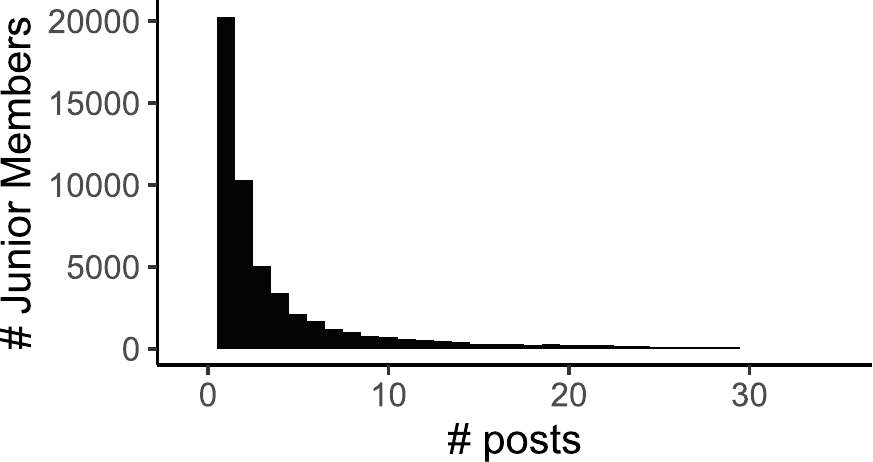}}
    \hfill
    \subfloat[\# Members vs. \# posts]{\includegraphics[width=.5\textwidth]{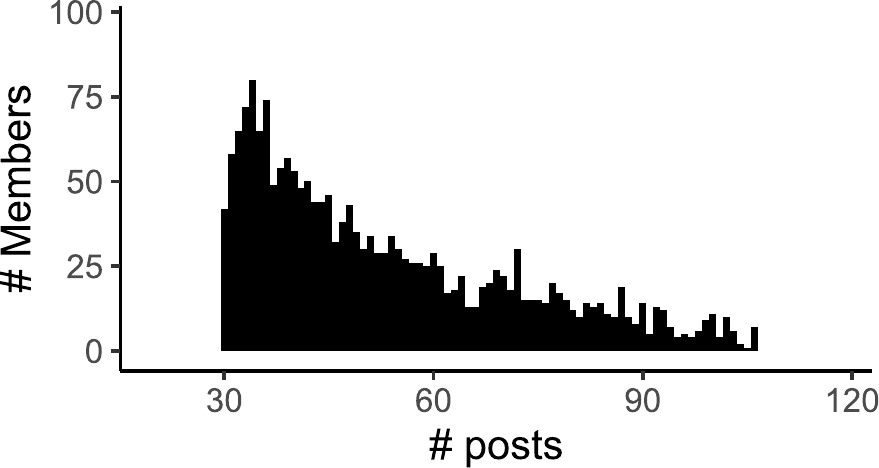}}
    \hfill
    \subfloat[\# Senior Members vs. \# posts]{\includegraphics[width=.5\textwidth]{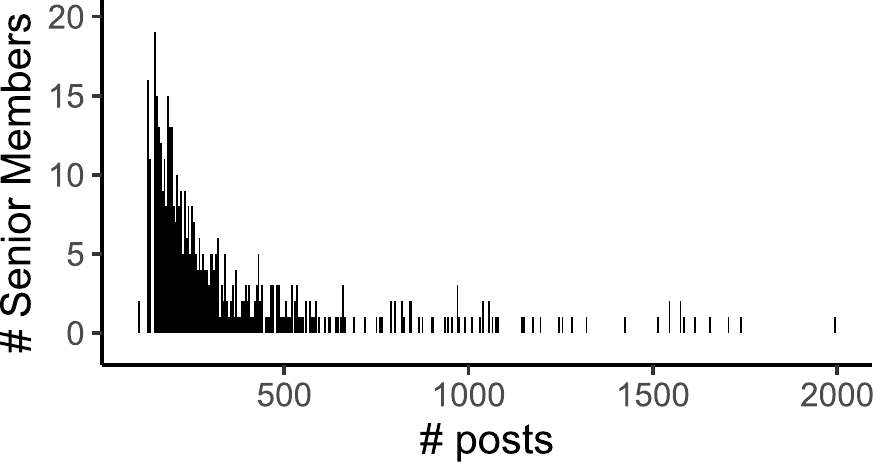}}
    \hfill
    \caption{\fix{Distribution of posts per user category after classification}}
    \label{fig:messagedistribution}
\end{figure*}

\textbf{Message distribution.}
Figure~\ref{fig:messagedistribution} shows the contributions of forum members in the Eclipse forums based on the memberships of our classification.
Even though the number of Junior Members is higher, they had posted a smaller quantity of posts, compared with the two other user categories.
In contrast, although the total number of Senior Members is lower than the other forum members, most Seniors have contributed in the forum multiple times.

\subsection{Online Appendix}
\label{ssec:appendix}

Our online appendix contains five data files used in this study: 
(i) 289,061 non-webmaster threads associated with the information of thread URL, message identity number, forum name and topic, thread title, name of initial-post user, user identity number, and member status; 
(ii) 216,864 links with the description of thread url, thread title, user's name, extracted link, the link with first directory, domain name, and top level domain; 
(iii) 2,170 contributors with the metric values described in Table~\ref{tab:variable};
(iv) 1,149 samples of the annotated threads to answer the types of contents; and
(v) 1,149 pairs of initial post and first response that are labeled based on the polarity and interaction sentiment.
The appendixes are available at \url{https://doi.org/10.5281/zenodo.4451766} and \url{https://github.com/yusufsn/EclipseForumData}.

\section{Methods}
\label{sec:method}

In this section, we describe our approach to answer the three research questions.

% \subsection{User participation (RQ$_1$)}
\subsection{\RqOne}
\label{sssec:approachPS1}

Before we study the participation of users' based on their membership statuses, how contributors participate to each system (i.e., gerrit, bugzilla, projects and forums) were analyzed.
We would like to understand how many of the contributors are using multiple systems, and whether or not they are forum users.

\textbf{Shape of contributors' participation.}
To show how contributors use the multiple systems we adopt a topological data analysis technique (i.e., TDA)~\citep{Lum13}.
 TDA is a topological technique to extract valuable information from high-dimensional data that is insensitive to metric, noisy, and incomplete without formulating an initial hypothesis.
This technique applies (algebraic) topology and computational geometry to create a shape of a high-dimensional dataset~\citep{Lum13}.

TDA has been applied in various fields such as an analysis of bacteria survival in soils~\citep{Ibekwe2014} and the performance of NBA players~\citep{Lum13}. 
Most recently in software engineering,~\cite{8305952} used TDA to distinguish characteristics of JavaScript npm libraries (such as licence usage, dependency usage and so on) and~\cite{Tantisuwankul2019CC} to analyze the implementation of such communication channels over open source software projects.
We use TDA to provide a visual representation that shows (i) the activity levels for each feature and (ii) show the contributor activity across systems, especially with respect to forums.

\begin{table}
    \centering
    \caption{Contributor metrics}
    \label{tab:variable}
    \begin{tabular}{llr}
        \toprule
        \textbf{metric} & \textbf{description} & \textbf{system} \\
        \midrule
        bug\_subm & \# of submitted bug reports & Bugzilla \\
        bug\_comm & \# of bug reports to which only are added comments & Bugzilla \\
        review\_subm & \# of submitted patches for review & Gerrit \\
        review\_comm & \# of reviews to which only are added comments & Gerrit \\
        thread & \# of threads participated in & Forums \\
        committer & \# of committer roles & Projects \\
        \bottomrule
    \end{tabular}
\end{table}

As shown in Table \ref{tab:variable}, we first prepare a contributor metrics data that will be used as features for the TDA mapper. 
This data contains 6 metrics that pertain to the different activities for each system from the 2,170 Gerrit contributors described in Section~\ref{ssec:cdc}.
We use the Knotter tool,\footnote{\url{https://github.com/rosinality/knotter}}
which  is  an  implementation  of  mapper  algorithm and the  t-Distributed  Stochastic  Neighbor  Embedding (t-SNE)\citep{vander2008visualizing}, a technique for dimensional reduction
and  clustering,  and  our  defined  features  as  the  filters  for the visualization construction.

For evaluation, we present the typologies of the four systems, highlighting the most active contributors.
To show this relationship in terms of their contributions to the forum, we will annotate the most active contributors of forums.

\textbf{Membership-based participation.}
To study how users participate in the forums based on their statuses, we divided the type of threads into two categories, (i) bug-related threads, and (ii) non-bug-related threads.
In this analysis, we targeted all 202,602 threads that were posted by the registered users (after excluding ``Eclipse User'' from the dataset), as shown in Table~\ref{tab:numberofrole}.
To distinguish the category of threads, we filtered the threads using the Eclipse bug-report URL as the keyword, that is, \url{bugs.eclipse.org}.
If the threads contain at least one keyword of the bug-report URL, the threads are categorized as bug-related threads, otherwise, non-bug-related threads.
Since we investigate how developers respond to bug-related and non-bug-related threads, the threads without any response were excluded.
After this process, we subsequently investigated 128,698 bug-related threads, and 25,052 non-bug-related threads.

For both categories of threads, we separated the targeted threads into three classes based on the status of the askers (initial poster), that is, Juniors' threads, Members' threads, and Seniors' threads.
In each thread, we then classified the response messages into the same classes as the askers to investigate the user types of the answers.
To reduce bias, the answers from the same authors who ask the questions in the threads were excluded.
We also removed the response messages with duplicate identification numbers of the answers in each thread.
Finally, we quantified the frequency of the answers grouped by their membership statuses.

In addition, to statistically validate our results, we applied Pearson's chi-squared test ($\chi^2$) \citep{pearson1992criterion} with the null hypothesis ``\textit{bug and non-bug related threads have independent participation by all members of different statuses.}'' The Pearson chi-square statistical test is commonly used for testing relationships between one or more categorical variables.

% \subsection{Eclipse forum contents (RQ$_2$)}
\subsection{\RqTwo}
\label{ssec:charofthreads}

This analysis includes the investigation of message patterns posted by the organization, the hottest topic of threads discussed in the forum, and the forum linkage to the external sources.

\textbf{Threads by organization.}
In the Eclipse community forums, a large number of \fix{posts} were posted by the organization
that deals with the servers and software that runs the eclipse.org site.\footnote{\url{https://wiki.eclipse.org/WebMaster}}
When the organization posts a message in the forums, they employ the same identity, that is ``Eclipse Webmaster''.
To understand how the Eclipse Webmaster manages its community forums, we used 542,997 messages collected in Section~\ref{ssec:ftc}.
\fix{Since we found there are a large number of duplicated messages in our collected dataset (e.g. the closure of forum listed in the topic name\footnote{\url{https://www.eclipse.org/forums/index.php/f/102/}}~\footnote{\url{https://www.eclipse.org/forums/index.php/sp/9957/}}), we removed them.}
After the removal of these duplicate messages, there were 934 messages with different contents. 
For these messages, we found some frequent patterns. 
Since the messages were unstructured and did not contain explicit keywords, we manually examined the contents of the 934 messages and classified them into 9 patterns.

\textbf{Forums and topic categories.}
To identify the forums and topics that are mostly discussed by the users, we quantitatively analyzed the messages based on their forum names and topic categories that are embedded in the collected threads.
We investigated 289,061 threads that were posted by the non-webmaster users, as presented in Table~\ref{tab:noofthreadeachstep}.

\textbf{Forums linkage.}
To understand how forum links to the external sources in the Eclipse ecosystem, we manually analyzed a representative sample of links.
The link targets are extracted from 289,061 threads posted by the users (see Table~\ref{tab:noofthreadeachstep}).
We first prepared a statistically representative sample from 216,864 collected links by computing a random sample of data with a confidence level of 95\% and a confidence interval of 5.\footnote{\label{surveysystem}\url{https://www.surveysystem.com/sscalc.htm}}
The calculation of the sample size yields 383 links.

To extract the usage of the link targets, we performed a manual labeling to all link targets from our sample (similar to \cite{Hata2019}). 
At this stage, the authors of this paper specified the code for categorizing the usage of the links.
The initial codes used to categorize the links were imported from the study by \cite{Hata2019}.
However, we dropped several labels because they are unrelated to our research.
We also combined and adjusted some codes to make them appropriate to our study.
Similar to other works in classifying the link targets~\citep{Hata2019}, four authors of this paper (the first, third, fifth, and the sixth authors) then coded the first 30 links from the representative sample independently using the designed codes.
The kappa agreement from the four raters is 0.81 or `almost perfect'~\citep{Viera:kappa:2005}.
Based on this encouraged agreement, the remaining link targets were then coded by the first author.

The codes used to characterize the link targets including the description are as follows:

\begin{itemize}
    \item \textit{404}: the link target cannot be accessed or missed
    \item \textit{bug report or Bugzilla}: a specific bug report or a Bugzilla top page
    \item \textit{other documentation}: documentation of a product or project in general except for API documentation
    \item \textit{personal or organization homepage}: a web page of an individual or organization
    \item \textit{product or project homepage}: a web page of a product or project
    \item \textit{API documentation}: specific documentation of an API component
    \item \textit{tutorial or article}: a tutorial or technical article without comments
    \item \textit{thread}: thread in the forums
    \item \textit{blog post}: informational website that displays postings by one or more individuals and usually has commenting section
    \item \textit{release}: a web page informing the release of new files, new versions of a software, new packages, etc.
    \item \textit{code}: a web page of a source code file
    \item \textit{book or research paper}: a web page of a book or entire book or academic paper
    \item \textit{licence}: licence of a software project
    \item \textit{other}: anything that does not fit the other labels, or a web page requiring sign-in. For example, a link that directs to a non-English web page~\url{http://www.vsa.de/marvin/dsl/mask/0.1} referenced in a forum content.~\footnote{\url{https://www.eclipse.org/forums/index.php/t/627844/}}
    
\end{itemize}

\textbf{Membership-based content.}
Our approach to categorize the content based on the user's statuses is through manual analysis of a representative sample.
Since this analysis only focus on the threads that were initially posted by the three types of registered users (i.e. Juniors, Members and Seniors), we excluded all threads that were initially posted by ``Eclipse Users''.
Similar to the procedure of forum linkage analysis, the statistically representative samples from the 202,602 threads were prepared.
The calculation of the sample size yields 379, 380, and 383 threads for Senior Member, Member, and Junior Member from a total of 202,602 threads.

To extract the content types, we conducted an interactive process of coding.
In this process, three authors of this paper (i.e. the first, fourth, and fifth authors) firstly discussed the initial coding guide from previous work~\citep{beyer2019kind}. 
To make the coding guide fit with our study, the initial coding guide were adjusted and the other codes were added for labeling the threads.
Similar to the approach from prior works~\citep{abdellatif2020challenges, Hata2019}, the authors subsequently coded 30 \fix{initial posts of a thread} independently for each type of askers in the sample using the designed labels.
The kappa scores for each status of askers were calculated to see the level of agreement between three authors.\footnote{\label{kappacalculator}\url{http://justusrandolph.net/kappa/}}
We gained 0.76 and 0.72 for both Juniors and Members which mean `acceptable', and 0.82 for Seniors which indicates `almost perfect'~\citep{Viera:kappa:2005}.
Based on these motivated agreement scores, the coding task for the remaining main posts from the sample were undertaken only by the first author.
The following terms list shows the labels used in our analysis to code the threads discussed including the descriptions:

\begin{enumerate}
    \item \textbf{Question-and-Answer threads}: if the thread is triggered by a question for asking a solution, reason, advice, clarification, or guidance even if the questions are unanswered, then we used the following coding guide:
    
    \begin{itemize}
        \item \textit{Errors}: questions about specific errors, including error messages, exceptions, the stack trace, etc.~\citep{beyer2019kind}.
        \item \textit{Review}: this category merges the categories Decision Help and Review~\citep{treude2011HowDoProgrammer}, the category Better Solution~\citep{beyer2014ManualCategorization}, and What~\citep{rosen2016mobiledev}, as well as How or Why something works~\citep{allamanis2013Why}. Questioners of these posts ask for better solutions or reviewing of their \textbf{code snippets}. Often, they also ask for best practice approaches or ask for help to make decisions, for instance, which API to select.
        \item \textit{Plans}: asking for the future plans or processes of the development.
        \item \textit{Learning}: this category merges the categories Learning a Language or Technology~\citep{allamanis2013Why} and Tutorials or Documentation~\citep{beyer2017analyzing}. In these posts, the \textbf{questioners ask for documentation, tutorials, or examples} to learn a tool or language. In contrast to the first category, they do not aim at asking for a solution or instructions on how to do something. Instead, they aim at asking for support to learn on their own.
        \item \textit{Usage}: this category subsumes questions of the types \textbf{How to implement something} and Way of using something~\citep{allamanis2013Why}, as well as the category How-to~\citep{beyer2014ManualCategorization, treude2011HowDoProgrammer}, and the Interaction of API classes~\citep{beyer2017analyzing}. Specific functionalities are mentioned in the question. The questioner is asking for concrete instructions.
        \item \textit{Versions}: this question category is equivalent to the categories Version~\citep{beyer2014ManualCategorization} and API Changes~\citep{beyer2017analyzing}. The questions arise due to the changes in an API or due to compatibility issues between different \textbf{versions} of an API. Specific versions are mentioned in the questions.
        \item \textit{Conceptual}: The posts of this category are equivalent to the category Conceptual~\citep{treude2011HowDoProgrammer} and subsumes the categories \textbf{Why...?} and \textbf{Is it possible...?}~\citep{beyer2014ManualCategorization}. Furthermore, it merges the categories \textbf{What}~\citep{rosen2016mobiledev} and \textbf{How/Why something works}~\citep{allamanis2013Why}. The posts consist of questions about the limitations of an API and API behaviour, as well as about understanding concepts, such as design patterns or architectural styles, and background information about some API functionality.
        \item \textit{Discrepancy}: the posts that are not questions of particular errors, but \textbf{no clue how to solve} questions~\citep{beyer2019kind}.
    \end{itemize}
    
    \item \textbf{Non-Question-and-Answer threads}: if the thread is triggered by a non-question post, we categorized the topics using the following labels:
    \begin{itemize}
        \item \textit{Misuse}: a post is identified out of the scope, unrelated to the community forum, or difficult to understand.
        \item \textit{Announcement}: a post provides an announcement from system or core developers about specific events (e.g. future updates, file release).
        \item \textit{Information}: a post provides a general information. It could be from anyone.
        \item \textit{Recruitment}: indicates an offer of a job vacancy or recruiting people.
        \item \textit{Test}: a post is used for a test.
        \item \textit{Other}: anything that does not fit the above labels, including posts that \fix{respond to a post} from different threads (the post is not an original question from the asker).
    \end{itemize}
\end{enumerate}

 To validate the results statistically, we then apply Pearson's chi-squared test ($\chi^2$) \citep{pearson1992criterion} with the null hypothesis ``\textit{the communicated content and the members of different statuses are independent.}"

% \subsection{Sentiment of interactions (RQ$_3$)}
\subsection{\RqThree}

Our approach to answer RQ$_3$ is through manual analysis of a representative sample to understand the sentiment of forum users based on their member statuses while sharing knowledge. 
To analyze the sentiment, we only focus on the \fix{initial posts of any threads} that were posted by three types of users (i.e., Juniors, Members and Seniors) and their first replies. 
This is because the first reply in each thread shows the tangible feeling of a user reaction in responding to the initial \fix{posts}.
Similar to Section~\ref{ssec:charofthreads}, we excluded all the initial posts from ``Eclipse User''. 
We then prepared statistically a representative sample for each users type with 95\% confidence interval.\textsuperscript{\ref{surveysystem}}
The calculation of the sample size yields 379, 380, and 383 pairs of initial post and first response for Senior Member, Member, and Junior Member, respectively.

To find the types of sentiment among users, we conducted an interactive process of coding. 
In this process, three authors of this paper first discussed the initial coding guide from previous work~\citep{kauffeld2012Meeting}. 
To make the coding guide fit with our study, the initial coding guides were adjusted. 
We (first, second, and sixth authors) then independently applied the adjusted coding rules on both initial \fix{posts} and the first responses of the first 30 samples of thread. 
We used the kappa score calculator to check the agreement level and find the score 0.78.\textsuperscript{\ref{kappacalculator}}
According to~\cite{Viera:kappa:2005}, this kappa agreement score is `acceptable'. 
Based on the agreement level, the coding tasks for the remaining threads from samples were undertaken by the three authors, which each author classified 383 different samples of threads.

Our sentiment analysis is composed of two different analyses.
First, we investigated the polarity more generally. Second, we performed a deeper analysis of the contents to characterize the social interaction.
The following terms list shows the labels used in our analysis to code the type of polarity and interactions while sharing knowledge through forums.

\begin{enumerate}
    \item \textbf{Polarity analysis}: To analyze the polarity of messages, we used the following three labels:
    \begin{itemize}
        \item \textit{Positive}: It includes posts that has positive feeling while reading and also include positive words (i.e., good, nice, working example, thanks a lot etc). For example: \textit{\begin{quote}
        Thanks, you have been a great help. It is enough for me.
        \end{quote}}
        \item \textit{Neutral}: It includes posts that has neutral feeling (e.g. started with positive then ended with negative talks) while reading or do not include any biasing words (i.e., positive, negative). For example: \textit{\begin{quote}
           See the Task List's view menu's ``Show UI Legend'' action for an explanation of the colors and a link to how to change them.
        \end{quote}}
        \item \textit{Negative}: It includes posts that has negative feeling while reading and also include negative words (i.e., error, bug, not working etc). For example:  \textit{\begin{quote}
            I've added some custom key binds in M8 and sometimes they don't work. They still show up in the preferences dialog, but they are not indicated on the menu items they are assigned to. Closing Eclipse and restarting fixed the problem. Anyone else seen this? I don't know of any repro steps at this point.
        \end{quote}}
    \end{itemize}
    
    \item \textbf{Interaction analysis}: To analyze the interactions between forum members in the threads, we used the following seven statements:
        \begin{itemize}
            \item \textit{Positive Procedural (PPc)}: It includes procedural statements that are goal oriented (i.e., pointing out or leading back to the topic), procedural suggestion (i.e, suggestions for further procedure), procedural questions (i.e., questions about further procedure), economical thinking (i.e, weighing costs/benefits), etc. 
            
            \item \textit{Positive Socioemotional (PS)}: It includes socioemotional talks like encouraging participation, agreeing suggestions, offering praise, etc.
            
            \item \textit{Positive Proactive (PPa)}: It includes proactive statements that discusses interesting ideas, planning actions, agreeing upon tasks to be carried out, etc.
            
            \item \textit {Neutral}: It includes statements that can't fall in the above categories or posts that are neutral in nature.
            
            \item \textit{Negative Procedural (NP)}: It includes procedural statements that talks more about failure of procedure, complaining the procedure (i.e., behaviour of API, code), irrelevant things, etc.
            
            \item \textit{Negative Socioemotional (NS)}: It includes socioemotional talks like criticizing, disparaging comments about others, etc.
            
            \item \textit{Negative Counteractive (NC)}: It includes counteractive statements that cover irrelevant problems, no action plan, terminating discussions, etc.
        \end{itemize}
\end{enumerate}

Similar to the statistical validation of the results in RQ$_1$ and RQ$_2$, Pearson's chi-squared test ($\chi^2$)~\citep{pearson1992criterion} is applied with the null hypothesis ``\textit{the sentiment of interactions and the members of different statuses are independent.}''

\section{Results}
\label{ssec:mainresults}
In this section, we present the answer to the research questions and describe the results.

% \subsection{User participation (RQ$_1$)}
\subsection{\RqOne}
\label{ssec:participationresult}

\begin{figure*}
    \centering
    \subfloat[Color activity by \texttt{threads} (Forums)]{\includegraphics[width=.46\textwidth]{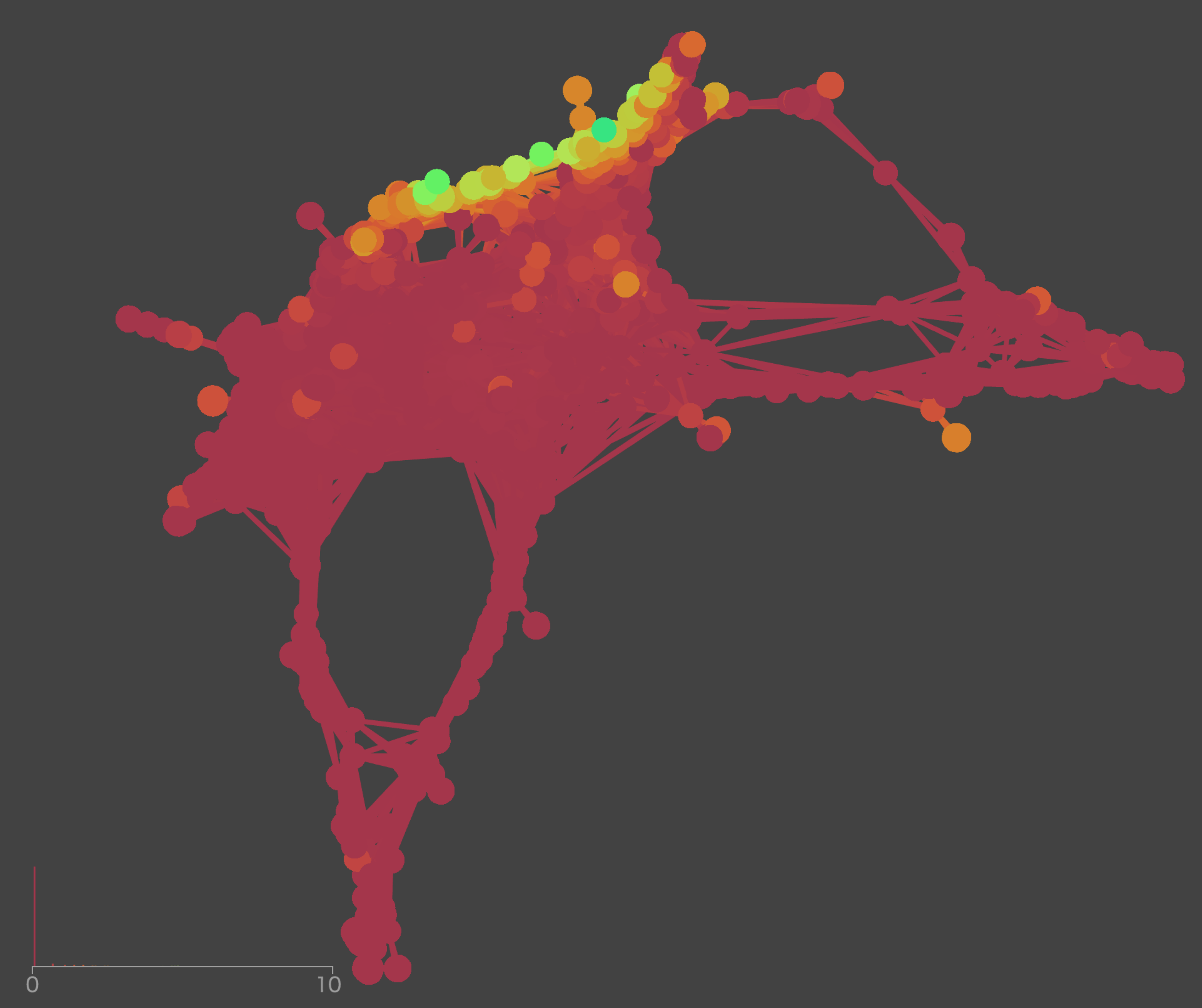}}
    \hfill
    \subfloat[Color activity by \texttt{bug\_subm} (Bugzilla)]{\includegraphics[width=.46\textwidth]{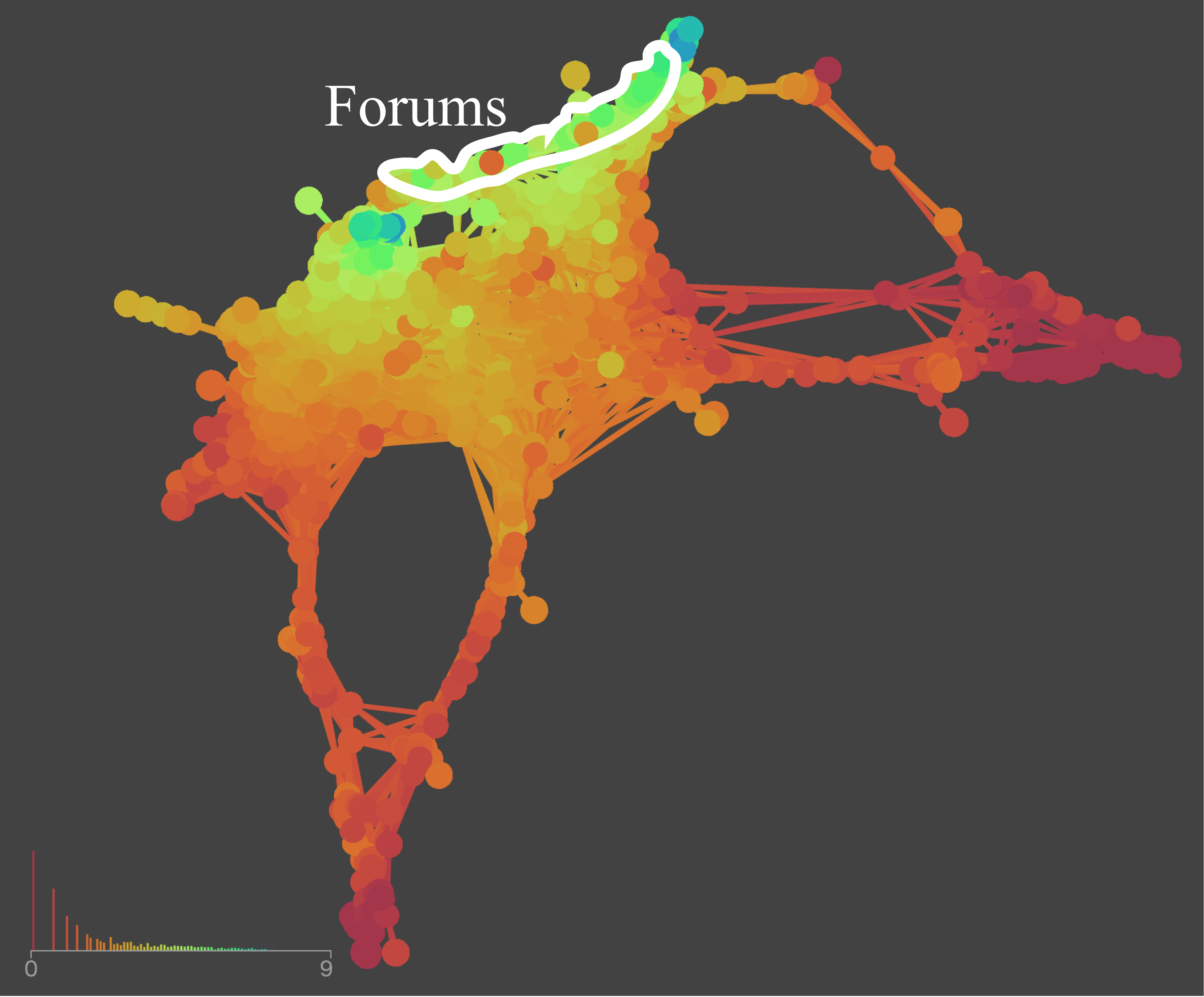}}
    \hfill
    \subfloat[Color activity by \texttt{review\_subm} (Gerrit)]{\includegraphics[width=.46\textwidth]{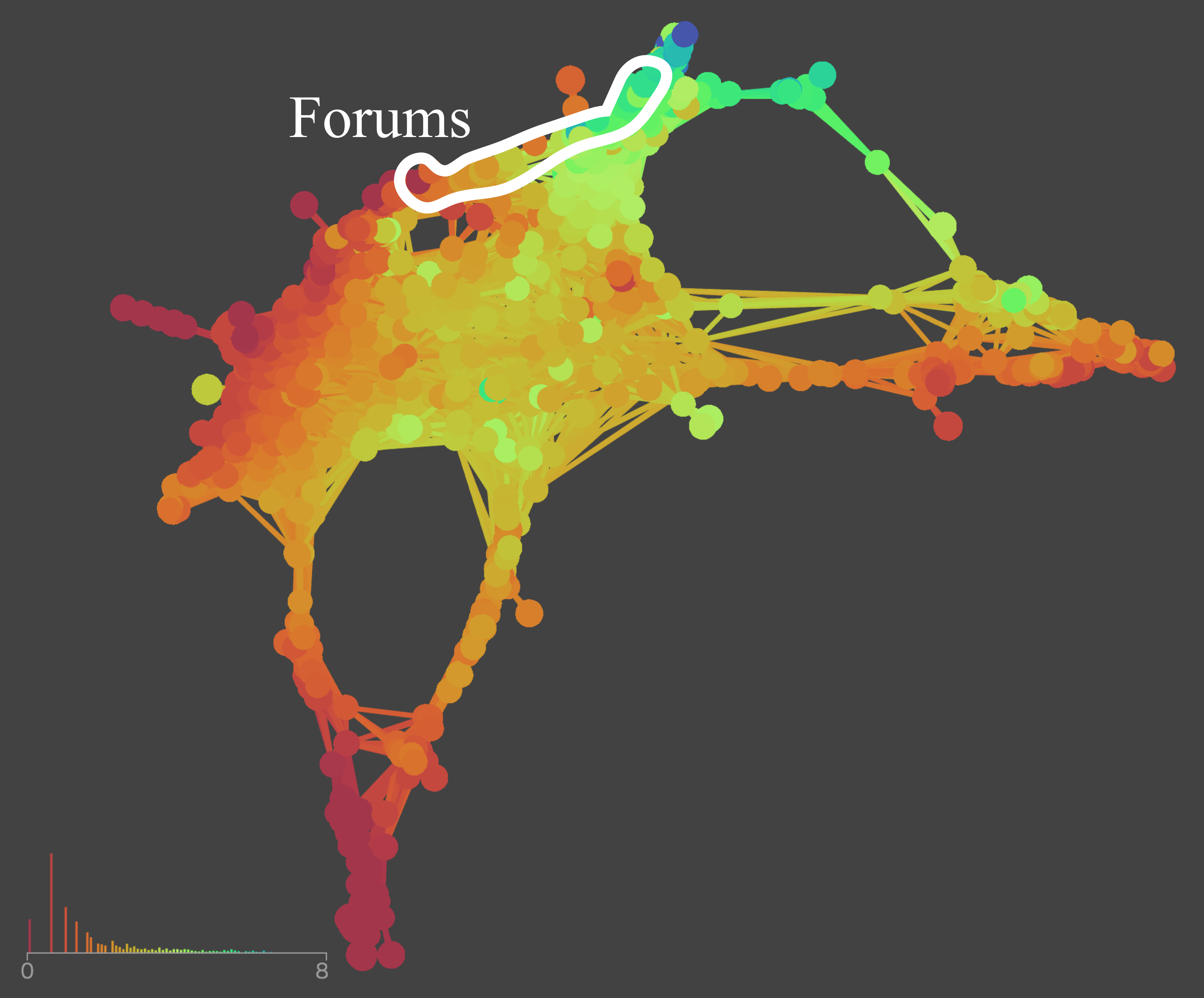}}
    \hfill
    \subfloat[Color activity by \texttt{committer} (Projects)]{\includegraphics[width=.46\textwidth]{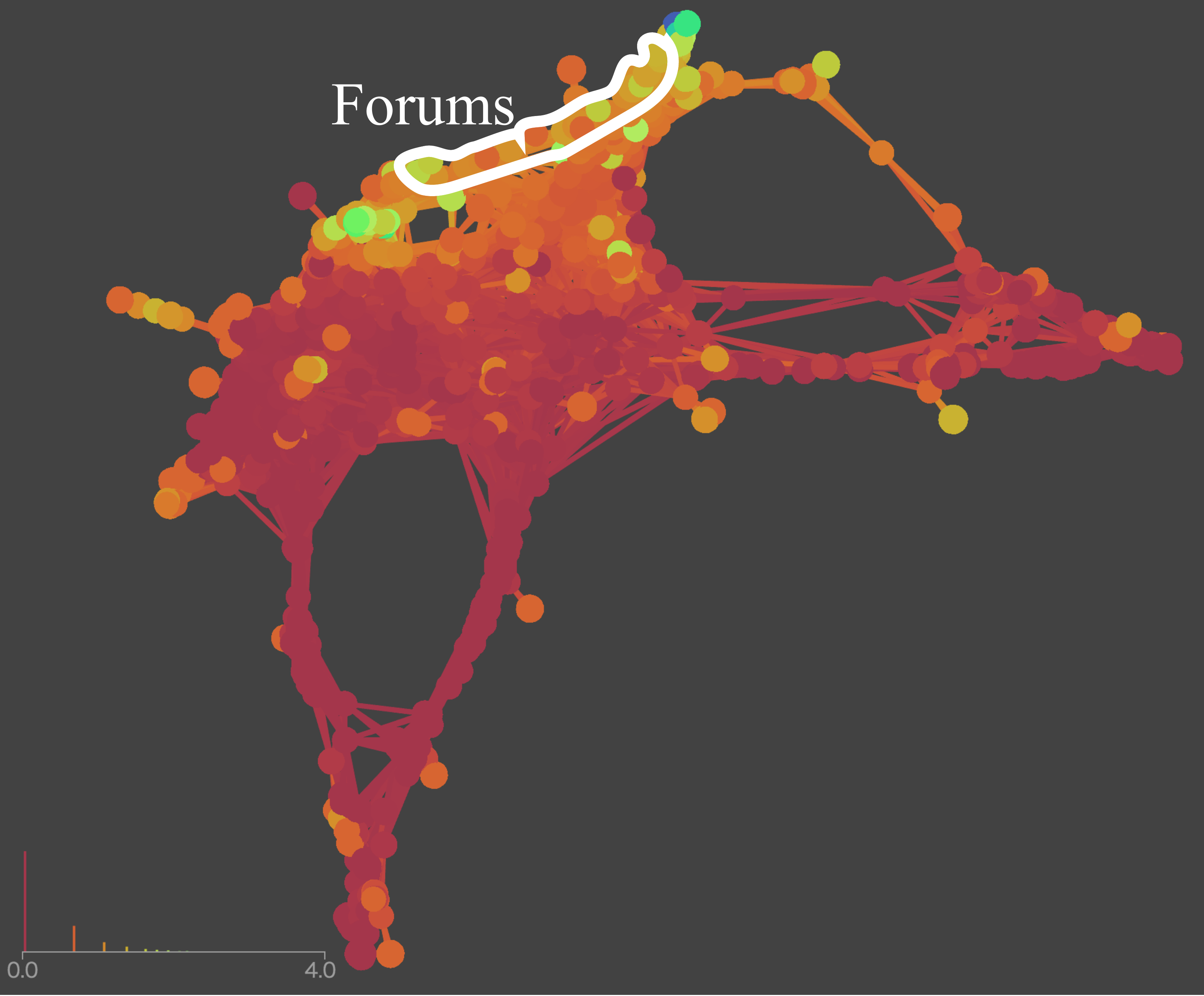}}
    \caption{The topology shows that Eclipse contributors with high activities in all systems (i.e., (a) Forums, (b) Bugzilla, (c) Gerrit, (d) Projects) are active in forums. Note that the metrics are taken from Table \ref{tab:variable}. We draw the white area manually to represent the contributors that actively participate in the forums.}
    \label{fig_sim}
\end{figure*}

Figure~\ref{fig_sim} shows the topology of active Eclipse contributors in all other systems.
Each node in the visualization represents similar sets of contributors.
In general, the map is read as follows:

\begin{enumerate}
    \item \textit{Topology cluster} - a cluster of nodes represent contributors that shared similar features (i.e., share similar contributions per systems).
    Hence, closely clustered nodes indicate these contributors share the same attributes.
    \item \textit{Topology color activity} - the color represent the density of each feature. Starting from red to blue, the red color (i.e., red=low activity) indicates a low activities of the feature, while green to blue color represented high activities (i.e., blue=high activity). For example, with the review comments feature, contributors that contributed many reviews to the Gerrit system were clustered in the green nodes, while those contributors with almost no activity are assigned red color.
\end{enumerate}

\textbf{Shape of contributors' participation.}
Figure~\ref{fig_sim} shows that the Eclipse contributors with high activities in all other systems are more active in the forums.
As shown in Figure~\ref{fig_sim}b and Figure~\ref{fig_sim}c, Eclipse contributors seem to work on Gerrit and Bugzilla systems to report bug and submit review.
This is because permission may be required for submission and committing of patches.
Hence it is not easily accessible.
Another interesting observation is that forums are used by the more socially active contributors. 
This is illustrated in Figure~\ref{fig_sim}d that those who are more active in Eclipse forums also participate more frequently in Eclipse projects as committers, as indicated in the white area of Figure~\ref{fig_sim}b and Figure~\ref{fig_sim}c.
This provides evidence that the forum is a source where expert knowledge can be shared.

\vspace{3mm}
\begin{tcolorbox}
    \textbf{Summary}: We find that active contributors are also active in forums, making it a source of expert knowledge for all systems.
\end{tcolorbox}

\begin{table}
    \centering
    \caption{Frequency of developers responding bug-related and non-bug-related threads }
    \label{tab:freqofdev_respondingthreads}
    \resizebox{\columnwidth}{!}{%
    \begin{tabular}{l|rrr|r|rrr|r} 
        \hline
        \multirow{2}{*}{\diagbox{Answerers}{Askers}} & \multicolumn{3}{c|}{Bug-related Threads (\%)} & \multicolumn{1}{c|}{sum} & \multicolumn{3}{c|}{Non-bug-related Threads (\%)} & \multicolumn{1}{c}{sum} \\ 
        \cline{2-4} \cline{6-8} 
        & \multicolumn{1}{c}{Juniors} & \multicolumn{1}{c}{Members} & \multicolumn{1}{c|}{Seniors} & 
        \multicolumn{1}{c|}{(\%)} & \multicolumn{1}{c}{Juniors} & \multicolumn{1}{c}{Members} & \multicolumn{1}{c|}{Seniors} &  
        \multicolumn{1}{c}{(\%)} \\ 
        \hline
        Juniors & 15.4 & 2.7 & 1.3 & 19.4 & 30.3 & 3.9 & 5.2 & 39.3 \\
        Members & 11.7 & 1.7 & 1.0 & 14.5 & 6.9 & 5.0 & 3.2 & 15.2 \\
        Seniors & 51.1 & 10.6 & 4.4 & 66.1 & 19.2 & 5.8 & 20.5 & 45.5 \\
        \hline
        \textbf{sum} & 78.8 & 15.1 & 6.8 & \textbf{100.0} & 56.4 & 14.7 & 28.9 & \textbf{100.0} \\
        \hline
    \end{tabular}%
    }
\end{table}

\textbf{Membership-based participation.}
As shown in Table~\ref{tab:freqofdev_respondingthreads}, Seniors seem to be the most active forum users in responding any category of threads, 66.1\% in bug-related threads and 45.5\% in non-bug-related threads, followed by Juniors and Members respectively.
In the category of non-bug-related threads, 
\fix{Juniors mostly responded to Juniors' initial posts, and Seniors mostly responded to Seniors' initial posts, while Members frequently responded to Juniors' initial posts.
In both categories of threads, Juniors and Members also replied to the Seniors.}
It indicates that the communication between forum users is not limited by the type of users.

For the statistical evaluation, we find that there is an association between the bug/non-bug related threads and the members of different status. Our null hypothesis on \textit{``bug and non-bug related threads have independent participation by all members of different statuses''} is rejected (i.e., p-value is
$<$ 0.001).

\vspace{3mm}
\begin{tcolorbox}
    \textbf{Summary}: 
   Compared to the Juniors and Members, Seniors are the most active forum users in responding any category of threads. For bug related threads, Seniors are contributing way more than the Juniors and Members. For non-bug related threads, Junior users responded more to Juniors' initial posts, and Senior users responded more to Seniors' initial posts.
\end{tcolorbox}

% \subsection{Eclipse forum contents (RQ$_2$)}
\subsection{\RqTwo}
\label{sec:discusstopic}

\textbf{Threads by organization.}
Table~\ref{tab:webmaster} shows the message pattern posted by the Eclipse webmaster. 
We see that the frequency of the reply and welcome \fix{messages} have surpassed other classes. 
The \texttt{Re:} messages are the most posted message patterns by the webmaster to indicate the responses to previous messages.
The Eclipse webmaster was also frequently sending a welcome message to the newly registered members in a particular community discussions group.
This type of message is used to provide a brief statement of the purpose of the newsgroups.

\begin{table}
    \centering
    \caption{Webmaster's message patterns}
    \label{tab:webmaster}
    \begin{tabular}{lp{8cm}r}
        \toprule
        \textbf{pattern} & \textbf{description (keywords)} & \textbf{\#} \\
        \midrule
        re: & a reply for a previous message (``\textit{Re:}'') & 538  \\
        welcome & a greeting to a newcomers (``\textit{Welcome to}'') & 270  \\
        closure & notification of closing threads (``\textit{closure}'', ``\textit{forum closed}'') & 25  \\
        test & a test message (``\textit{test}'', empty message) & 14  \\
        archive & notification of archiving threads (``\textit{been archived}'') & 13  \\
        guideline & rules, intention, or suggestions for users (``\textit{posting guidelines}'', ``\textit{please read before posting}'') & 6  \\
        announce & notification of specific events or internal conditions (``\textit{outage}'', ``\textit{we're hiring!}'', ``\textit{available}'', ``\textit{shutting down}'') & 5  \\
        move & notification of moving threads (``\textit{this group has moved to}'', ``\textit{forum move}'') & 4  \\
        other & a message that does not fit the above & 59  \\
        \midrule
        \textbf{sum} & & \textbf{934} \\
        \bottomrule
    \end{tabular}
\end{table}

Despite the occurrences of the closure messages are not as frequent as the top 2 patterns, but they were posted by the organization occasionally to inform the contributors that the related forums had been closed.
The other pattern of messages that have a number of messages posted in the forum are test and archive, as many as 14 and 13 message contents, followed by guideline, announce and move, with 6, 5 and 4 messages, respectively. 
Finally, the prevalence of the pattern ``other'' in the findings is an indicator of the various message patterns in the threads posted by the Eclipse Webmaster.

\textbf{Forum and topic categories.}
Table~\ref{tab:toptentopicsinforum} shows that the Eclipse platform is the most common topic discussed by Eclipse users.
This may imply that the threads are more generic, rather than being very specific.
The next topic is followed by BIRT, i.e., an open source software project to create data visualizations and reports.
Both topics are in the forum of Eclipse projects.
In our statistical records, Newcomers has become the third most topic in the Eclipse forum. 
This conversation subject is used dominantly by those who are new in the Eclipse community.
The other prevalent topics are EMF of Modeling, C/C++ IDE (CDT) and Java Development Tools (JDT) of Language IDEs, to complete the top six topics discussed in the Eclipse community forums.

\begin{table}
    \centering
    \caption{Top 10 forum and topic categories in the Eclipse community forums}
    \label{tab:toptentopicsinforum}
    \begin{tabular}{llr}
        \toprule
        \textbf{forum} & \textbf{topic} & \textbf{\# threads}  \\
        \midrule
        Eclipse Projects & Eclipse Platform & 31,795 \\
        & BIRT & 29,019 \\
        Newcomers & Newcomers & 23,054    \\
        Modeling & EMF & 16,171    \\
        Language IDEs & C / C++ IDE (CDT) & 13,449 \\
        & Java Development Tools (JDT) & 13,193  \\
        Eclipse Projects & Standard Widget Toolkit (SWT) & 11,307    \\
        & Rich Client Platform (RCP) & 10,842    \\
        Modeling & TMF (Xtext) & 9,909  \\
        & GMF (Graphical Modeling Framework) & 8,484    \\
        \bottomrule
    \end{tabular}
\end{table}

\begin{table}
    \centering
    \caption{Frequency of link target types in our sample}
    \label{tab:freqoflinktargettypes}
    \begin{tabular}{p{1.5cm}lrr}
        \toprule
        \textbf{availability} & \textbf{target} & \textbf{\# links} & \textbf{(\%)} \\
        \midrule
        \rowcolor{gray}
        \multicolumn{3}{l}{available} & (72\%)  \\
        & bug report/Bugzilla   & 58 & (15\%) \\
        & other documentation   & 41 & (11\%) \\
        & personal/organizational homepage & 38 & (10\%) \\
        & product/project homepage & 28 & (7\%) \\
        & API documentation     & 20 & (5\%) \\
        & tutorial or article   & 16 & (4\%) \\
        & thread                & 15 & (4\%) \\
        & blog post             & 10 & (3\%) \\
        & release               & 8 & (2\%) \\
        & code                  & 4 & (1\%) \\
        & book or research paper & 1 & (0\%) \\
        & licence               & 1 & (0\%) \\
        & other                 & 34 & (9\%) \\
        \rowcolor{gray}
        \multicolumn{3}{l}{not available} & (28\%)  \\
        & 404                   & 109 & (28\%) \\
        \midrule
        \textbf{sum} & & \textbf{383} & \textbf{(100\%)} \\
        \bottomrule
    \end{tabular}
\end{table}

\textbf{Forums linkage.}
Table~\ref{tab:freqoflinktargettypes} shows the result of our qualitative analysis.
From Table~\ref{tab:freqoflinktargettypes}, we see that a bug-related link target from online issue trackers, such as Bugzilla, is the most links inserted in the messages as a supplement to the answer, accounting for 15\%.
It indicates that facing problems or finding a software defect was frequently reported by the users in the forum.
Other documentations are also prevalent, nearly the same as personal or organizational homepages, accounting for 11\% and 10\% respectively.
The common use of the label `other' in the link target types represents the heterogeneous of links present in the Eclipse forum. 
Lastly, we also see that 28\% of link targets referenced by the Eclipse forum members are currently inaccessible (i.e. 404).

\begin{table*}
    \centering
    \caption{Frequently referenced domains in the Eclipse community forums}
    \label{tab:top10mostdomain}
    \begin{adjustbox}{width=\textwidth}
    \begin{tabular}{llp{6cm}rr}
        \toprule
        \textbf{domain} & & \textbf{description} & \textbf{\# links}   \\
        \midrule
        www.eclipse.org & & & 39,917  \\
         & /forums/ & Eclipse Community Forums & & (5,723)  \\
         & /modeling/ & Eclipse Modeling Project & & (4,076)  \\
         & /emf/ & Eclipse Modeling Framework (EMF) (moved to under the above /modeling/) & & (3,119)\\
         & /birt/ & Eclipse BIRT (Business Intelligence and Reporting Tools) Project & & (2,391)  \\
        bugs.eclipse.org & & Eclipse Bugzilla & 30,135  \\ 
        wiki.eclipse.org & & Eclipse wiki pages & 22,920    \\
        download.eclipse.org & & download page for Eclipse product & 9,343    \\
        dev.eclipse.org & & (currently not available or redirected to wiki.eclipse.org/Development\_Resources) & 9,297 \\
         & /viewcvs/ & Eclipse CVS repositories (not available) & & (4,528) \\
         & /newslists/ & Eclipse news bulletin (not available) & & (2,533) \\
        xtext.itemis.com & & Xtext framework for programming language development & 3,459    \\
        www.w3.org & & international standard organization of world wide web & 3,323  \\
        github.com & & web-based hosting service for version control using Git & 3,298  \\
        twitter.com & & an online social networking site & 3,096 \\
        git.eclipse.org & & Eclipse Git repositories & 2,123 \\
        \bottomrule
    \end{tabular}
    \end{adjustbox}
\end{table*}

\textbf{Frequently linked resources.}
Table~\ref{tab:top10mostdomain} shows that the top 10 most prevalent referenced domains from the collected 216,864 links.
We see that the Eclipse organizational homepage is the most popular link to refer to, which is separated into four most common directories: Forums, and some Eclipse projects (Modeling, EMF, and BIRT).
Bugzilla, a web-based general-purpose bug tracker, is the second most prevalent domain to be referenced, followed by the wiki pages and the download pages of Eclipse.
Reporting newly discovered issues, or referencing the existing bugs in the Bugzilla is very common amongst the Eclipse forum discussions.
The domain of \texttt{dev.eclipse.org}, which is currently not available, was also frequently referenced, especially for CVS repositories and news bulletins.
We also found that the link resources from outside the Eclipse organization are preferred by forum users to refer to.
The most communal external links posted in the threads are from \texttt{xtext.itemis.com}, i.e. the originator of the Xtext framework, \texttt{www.w3.org}, i.e. the world wide web standard organization, \texttt{github.com}, i.e. distributed version-control platform, and \texttt{twitter.com}, i.e. online news and social network.
In addition, the Eclipse Git repositories website (\texttt{git.eclipse.org}) remains hot in the forum threads to complement the top 10 referenced domains.

\vspace{3mm}
\begin{tcolorbox}
    \textbf{Summary}: 
    Specific projects of \texttt{Eclipse Platfrom} and \texttt{BIRT}, and the forums for \texttt{Newcomers} are most active categories.
    We found that referencing to other resources, like bug reports, documentation, etc., is common in forum discussions, which indicates that forums are essential platform for linking various resources in Eclipse ecosystem.
\end{tcolorbox}

% ==============Table knowledge type==========
\begin{table}
    \centering
    \caption{Frequency of content types in our sample (gray color represents question-and-answer threads, white color represents non-question-and-answer threads)}
    \label{tab:freqofthreadpurpose}
    \resizebox{\columnwidth}{!}{%
    \begin{tabular}{l|rr|rr|rr|rr} 
        \hline
        \multicolumn{1}{c|}{\multirow{2}{*}{\textbf{Topic}}} & \multicolumn{2}{c|}{\textbf{Juniors}} & \multicolumn{2}{c|}{\textbf{Members}} & \multicolumn{2}{c|}{\textbf{Seniors}} & \multicolumn{2}{c}{\textbf{Total}} \\ 
        \cline{2-9}
        \multicolumn{1}{c|}{} & \multicolumn{1}{c}{\textbf{\#}} & \multicolumn{1}{c|}{\textbf{\%}} & \multicolumn{1}{c}{\textbf{\#}} & \multicolumn{1}{c|}{\textbf{\%}} & \multicolumn{1}{c}{\textbf{\#}} & \multicolumn{1}{c|}{\textbf{\%}} & \multicolumn{1}{c}{\textbf{\#}} & \multicolumn{1}{c}{\textbf{\%}} \\ 
        \hline
        \rowcolor{gray}
        Discrepancy & 153 & (39.9\%) & 114 & (30.0\%) & 58 & (15.3\%) & 325 & (28.5\%) \\
        \rowcolor{gray}
        Review & 51 & (13.3\%) & 63 & (16.6\%) & 39 & (10.2\%) & 153 & (13.4\%) \\
        \rowcolor{gray}
        Conceptual & 31 & (8.1\%) & 50 & (13.1\%) & 62 & (16.2\%) & 143 & (12.4\%) \\
        \rowcolor{gray}
        Usage & 45 & (11.7\%) & 36 & (9.4\%) & 44 & (11.6\%) & 125 & (10.9\%) \\
        \rowcolor{gray}
        Errors & 49 & (12.8\%) & 41 & (10.7\%) & 31 & (8.2\%) & 121 & (10.6\%) \\
        Information & 6 & (1.6\%) & 17 & (4.4\%) & 45 & (11.9\%) & 68 & (6.0\%) \\
        \rowcolor{gray}
        Learning & 19 & (5.0\%) & 19 & (5.0\%) & 13 & (3.4\%) & 51 & (4.4\%) \\
        \rowcolor{gray}
        Versions & 9 & (2.3\%) & 22 & (5.7\%) & 18 & (4.7\%) & 49 & (4.3\%) \\
        Announcement & 3 & (0.8\%) & 5 & (1.3\%) & 22 & (5.7\%) & 30 & (2.6\%) \\
        \rowcolor{gray}
        Plans & 8 & (2.1\%) & 4 & (1.0\%) & 9 & (2.3\%) & 21 & (1.8\%) \\
        Misuse & 1 & (0.3\%) & 1 & (0.3\%) & 6 & (1.6\%) & 8 & (0.7\%) \\
        Recruitment & 2 & (0.5\%) & 1 & (0.3\%) & 2 & (0.5\%) & 5 & (0.4\%) \\
        Test & 0 & (0.0\%) & 2 & (0.5\%) & 3 & (0.8\%) & 5 & (0.4\%) \\
        Other & 6 & (1.6\%) & 5 & (1.3\%) & 27 & (7.0\%) & 38 & (3.3\%) \\
        \hline
        \textbf{sum} & \textbf{383} & \multicolumn{1}{l|}{\textbf{(100\%)}} & \textbf{380} & \multicolumn{1}{l|}{\textbf{(100\%)}} & \textbf{379} & \textbf{(100\%)} & \textbf{1,142} & \textbf{(100\%)} \\ 
        \hline
    \end{tabular}%
    }
\end{table}

\textbf{Membership-based content}.
The results of our manual classification show that the Eclipse forum is very much similar to other community-based question-and-answer sites, dominated by the users that share problems through questions to get the answers from the community.
As shown in Table~\ref{tab:freqofthreadpurpose}, discrepancy is the most common type of problems shared by the forum members, especially from Junior Members.
Review, conceptual, usage, and errors complete the top five types of content that dominate in the Eclipse forum shared by all forum users, which accounts for more than 10\% for each type.
For the non-question-and-answer threads, information and announcement are two types of content that were most frequently shared between the forum members.

Our other findings on the content types that shared amongst the forum members at least hint that the forum is not only utilized for asking solutions or responses from other users, but also to announce some specific events or updates, distributing general information, offering some job vacancies, or even just testing the forum as well.

% =================== stacked bar ===========================

\pgfplotsset{compat=1.14}
%\definecolor{findOptimalPartition}{HTML}{696969}
\definecolor{disc}{HTML}{7B7B7B}%{99CCFF}
\definecolor{conc}{HTML}{AAAAAA}%{FFFFAA}
\definecolor{inf}{HTML}{CCCCCC}%{FFCCCB}
\definecolor{anno}{HTML}{EFEFEF}

\pgfplotstableread[col sep=comma,header=true]{% added header row
users, 1, 2, 3, 4
Juniors, 39.9, 8.1, 1.6, 0.8
Members, 30, 13.1, 4.4, 1.3
Seniors, 15.3, 16.2, 11.9, 5.7
}\threaddata

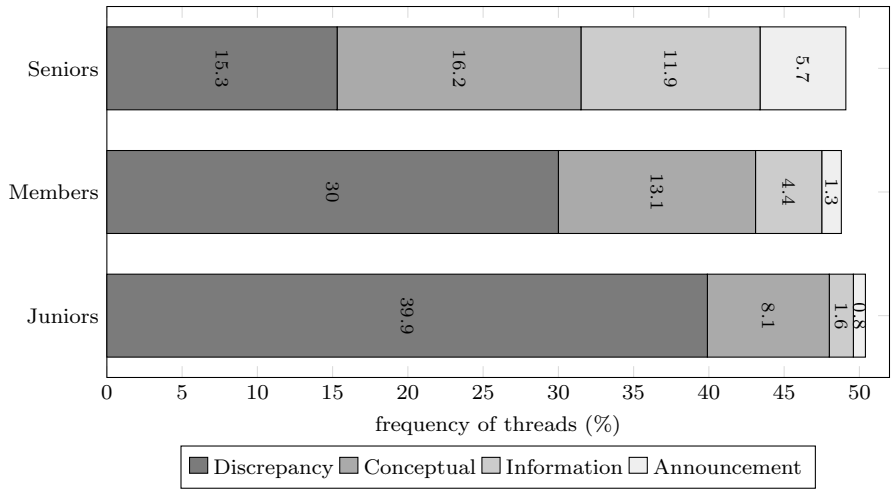
\begin{figure}
    \center
    \begin{tikzpicture}
        \begin{axis}[
            xbar stacked,   % Stacked horizontal bars
            xmin=0,         % Start x axis at 0
            ytick=data,
            ymin=-0.5,
            ymax=2.5,
            xmax = 52,
            xlabel={frequency of threads (\%)},
            height=6.5cm,
            width=\columnwidth,
            bar width=1.1cm,
            yticklabels from table={\threaddata}{users}, 
            legend style={
                at={(0.5,-0.3)},
                anchor=south,
                legend columns=-1
            },
            nodes near coords,
            % every node near coord/.append style={font=\scriptsize, anchor=north,inner ysep=1pt},
            every node near coord/.append style={font=\scriptsize, rotate=270},
            nodes near coords align=center
        ]
        \addplot [fill=disc] table [x=1, meta=users, y expr=\coordindex] {\threaddata};   % "First" column against the data index
        \addplot [fill=conc] table [x=2, meta=users, y expr=\coordindex] {\threaddata};
        \addplot [fill=inf] table [x=3, meta=users, y expr=\coordindex] {\threaddata};
        \addplot [fill=anno] table [x=4, meta=users, y expr=\coordindex] {\threaddata};
        
        \legend{\strut Discrepancy, \strut Conceptual, \strut Information, \strut Announcement}
        \end{axis}
    \end{tikzpicture}
    \caption{Four content types communicated in the forums}
    \label{fig:highlightpurposes}
\end{figure}

Our findings also highlight that there are several distinguished types of content that were communicated between the members based on their statuses.
As illustrated in Figure~\ref{fig:highlightpurposes}, discrepancy is the most common type that was shared by Junior Members.
However, the number of this type reduces once the status of the user levels up into higher status.
Conversely, the quantity of conceptual contents increases inline with the changes of the user statuses, from the lowest level of membership (Juniors) to the highest one (Seniors).
This shows that the higher the membership status of a user, the higher the level of the problems discussed in the forums. 
Furthermore, Senior Members posted general information and announcement more often than the other types of users.
The information shared in the forums is commonly general events, such as a seminar, presentation or competition to motivate users' participation.
While the announcement describes specific events related to products, for example, file release, future updates, etc. 
This result at least hints that most of the Senior Members are core members of a software development company who act as sources of information 
and convey the information to inspire and motivate other people~\citep{yunwen2003toward}.

For the statistical evaluation, we find that there is an association between the communicated content and the members of different status. Our null hypothesis on \textit{``the communicated content and the members of different statuses are independent.''} is rejected (i.e., p-value is $<$ 0.001).

\vspace{3mm}
\begin{tcolorbox}
    \textbf{Summary}: The Eclipse forums are dominated by question-and-answer threads, especially discrepancies, most of which were posted by Junior Members. 
    Furthermore, the status of users relates to the topics discussed in the forums. 
    The higher the status of users, the more conceptual the content type they communicate in forums.
\end{tcolorbox}

% \subsection{Sentiment of interactions (RQ$_3$)}
\subsection{\RqThree}
\label{sec:sentimentanalysis}

\begin{table}
    \centering
    \caption{Polarity of communication among developers}
    \label{tab:polarityofQA}
    \begin{tabular}{l|l|rrrr|r} 
        \hline
        \multicolumn{1}{l|}{\textbf{users'}} & \multicolumn{1}{c|}{\multirow{2}{*}{\textbf{main post}}} & \multicolumn{4}{c|}{\textbf{first response (\%)}} & \multicolumn{1}{c}{\multirow{2}{*}{\textbf{sum (\%)}}}  \\ 
        \cline{3-6}
        \multicolumn{1}{l|}{\textbf{type}} & \multicolumn{1}{c|}{} & \multicolumn{1}{c}{positive} & \multicolumn{1}{c}{neutral} & \multicolumn{1}{c}{negative} & \multicolumn{1}{c|}{no response} & \multicolumn{1}{c}{} \\ 
        \hline
        \multirow{3}{*}{Juniors} & positive & 24.3 & 17.5 & 5.7 & 13.8 & \textbf{61.4} \\
        & neutral & 2.1 & 3.1 & 0.8 & 4.4 & \textbf{10.4} \\
        & negative & 6.0 & 8.4 & 7.3 & 6.5 & \textbf{28.2} \\
        \hline
        \multicolumn{2}{l|}{\textbf{sum}} & \textbf{32.4} & \textbf{29.0} & \textbf{13.8} & \textbf{24.8} & \textbf{100.0} \\
        \hline
        \multirow{3}{*}{Members} & positive & 36.6 & 4.7 & 3.9 & 16.6 & \textbf{61.8} \\
        & neutral & 2.6 & 0.5 & 0.5 & 5.3 & \textbf{8.9} \\
        & negative & 18.4 & 2.1 & 2.1 & 6.6 & \textbf{29.2} \\
        \hline
        \multicolumn{2}{l|}{\textbf{sum}} & \textbf{57.6} & \textbf{7.4} & \textbf{6.6} & \textbf{28.4} & \textbf{100.0} \\
        \hline
        \multirow{3}{*}{Seniors} & positive & 14.8 & 6.3 & 3.7 & 19.8 & \textbf{44.6} \\
        & neutral & 5.0 & 4.2 & 2.4 & 13.2 & \textbf{24.8} \\
        & negative & 10.8 & 4.5 & 4.5 & 10.8 & \textbf{30.6} \\ 
        \hline
        \multicolumn{2}{l|}{\textbf{sum}} & \textbf{30.6} & \textbf{15.0} & \textbf{10.6} & \textbf{43.8} & \textbf{100.0} \\
        \hline
    \end{tabular}
\end{table}

\textbf{Polarity detection.}
Table~\ref{tab:polarityofQA} shows the polarity results of main posts and first responses among developers. 
We found that Juniors and Members post more positive messages than Seniors.
On the other hand, compared with the other types of developers, neutral and negative messages were mostly posted by Senior developers.
\fix{
We also observed that overall positive posts received more positive replies.
}

\begin{figure*}
    \centering
    \subfloat[Juniors thread\label{fig:Juniorinteraction}]{
        \includegraphics[width=.47\linewidth]{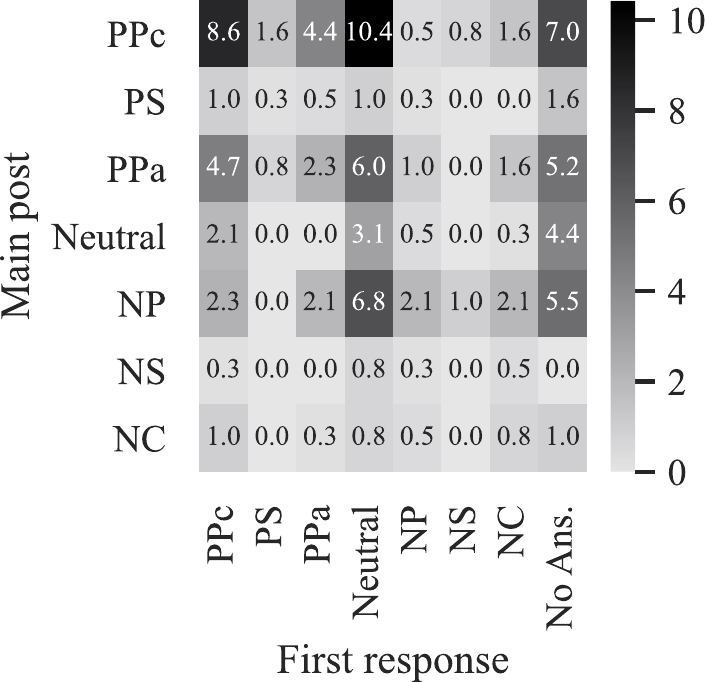}}
    \subfloat[Members thread\label{fig:memberinteraction}]{
        \includegraphics[width=.47\linewidth]{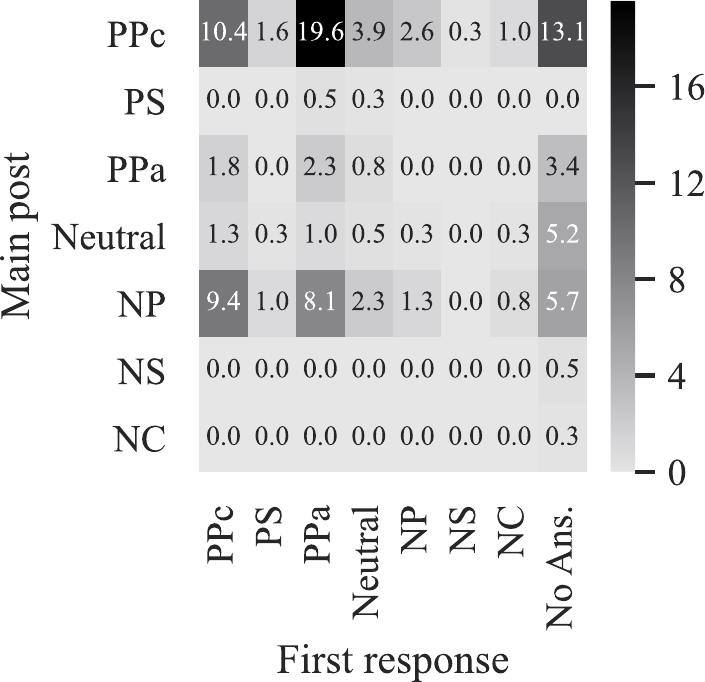}}
    \hfill
    \subfloat[Seniors thread\label{fig:seniorinteraction}]{
        \includegraphics[width=.47\linewidth]{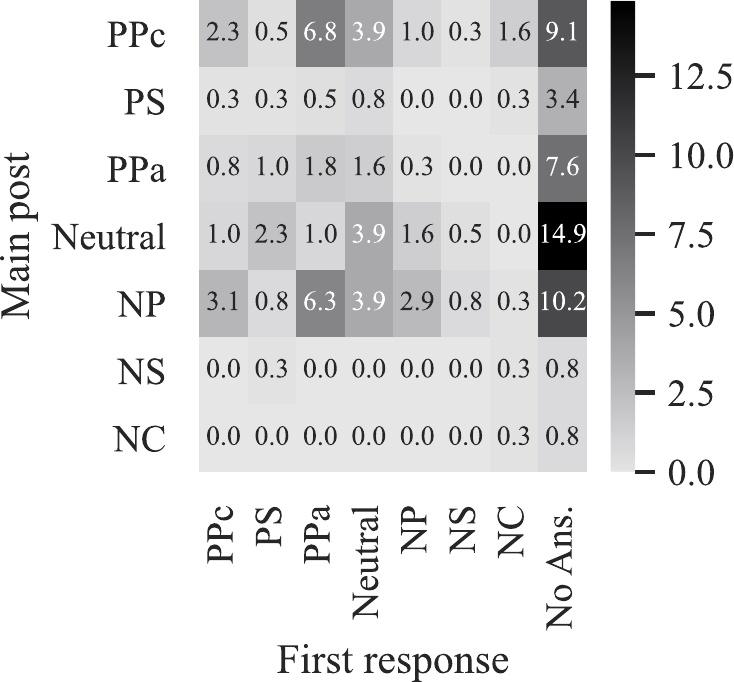}}
    \subfloat[Combination of all user types\label{fig:allinteraction}]{
     \includegraphics[width=.47\linewidth]{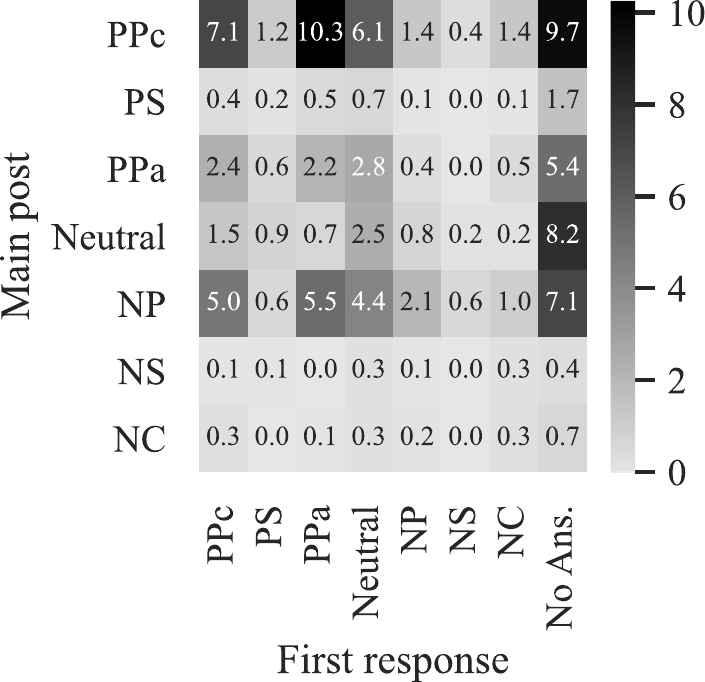}}
        
    \caption{Nature of first responses in the interaction between forum members. The color scale represents the frequency of the first responses in percentage based on total number of analyzed samples. The darker the area in the heatmap, the more frequent the first responses in the threads.}
    \label{fig:natureoffirstresponse}
\end{figure*}

\textbf{Social interaction.}
Figure~\ref{fig:natureoffirstresponse} shows the nature of first responses in the interaction among different type of developers. 
From Figure~\ref{fig:Juniorinteraction} and~\ref{fig:memberinteraction}, we observed that social interaction by Juniors and Members are overall positive. 
We found that Juniors and Members posted the initial \fix{posts} in a more positive procedural way. 
For this positive procedural interaction, Juniors receive more positive procedural and neutral responses while Members receive both positive proactive and no answer as responses.

In contrast with the other types of developers, Figure~\ref{fig:seniorinteraction} shows that Seniors post positive procedural, neutral and negative procedural initial posts. 
Complement to {RQ$_2$}, the neutral \fix{posts} mostly include an announcement related to a release of a new product, and general information such as advertisement, seminar, etc. 
In addition, Seniors' main posts did not receive responses in a large number of threads. 
This hints that Seniors seem to post more complex questions than Juniors and Members that may be difficult to answer.
If we combine all types of developers together as illustrated in Figure~\ref{fig:allinteraction}, we observed that the social interaction among developers are generally similar to the polarity.
\fix{
The result shows that, positive procedural posts received more positive proactive responses. 
}
Furthermore, posting a message in a positive procedural way has higher chances of getting positive replies.

For the statistical evaluation, we find that there is an association between the sentiment of interactions and the members of different status. Our null hypothesis on \textit{``the sentiment of interactions and the members of different statuses are independent."} is rejected (i.e., p-value is $<$ 0.001).

\vspace{3mm}
\begin{tcolorbox}
    \textbf{Summary}: We found that sentiment among developers are inconsistent while knowledge sharing. Interaction among developers shows that positive procedural initial posts overall have more chances to receive positive first responses.  
\end{tcolorbox}

\section{Recommendations}
\label{sec:recommendation}
Based on our study results, we provide three sets of recommendation for forum users, software development projects, and researchers.

For forum users, the recommendation regarding the threads and referencing links are as follows:
\begin{itemize}
    
    \item \textit{Join forum discussions},
    as forums play a role as the center of knowledge sharing platform. The users will not only receive knowledge from community, but also from external sources.

    \item \textit{Identify appropriate forum topic}.
    As shown in Table~\ref{tab:freqofthreadpurpose}, although the percentage of misuse contents is not high, some amount of users posted their messages in the incorrect discussion groups.
    Sharing problems in a compatible topic category would maximize the probability to get responses and solutions. Otherwise, the questions might be marked out of scope.
    
    \item \textit{Share knowledge in positive manner}.
    As shown in the results, users seem to respond positively to a positive and procedural message.
    Since the clarity of knowledge shared in the forums might increase the chances to get responses, describing the problems in a procedural way is essential.

\end{itemize}

One recommendation to software development projects is as follows:
\begin{itemize}
    \item \textit{Consider preparing project-specific forums}.
    As observed in the study, Eclipse forums is not only limited to question-and-answer based content, but also enable developers to provide Eclipse community-related information or announcement.
    Since the information or announcement that specifically relates to the software projects is important to share amongst the community, providing a project-specific forum is necessary.
    It has been reported that one of the keys to making OSS projects sustainable is to reduce the contribution cost by preparing the development environment~\citep{10.1109/CHASE.2015.9}.
    GitHub Discussions, a new feature of GitHub for asking questions or discussing topics outside of specific Issues or Pull Requests,\footnote{\url{https://docs.github.com/en/discussions}} could be an option for projects hosted on GitHub~\citep{2021arXiv210205230H}.
\end{itemize}

For researchers, we can consider future works with the following possible challenges:
\begin{itemize}
    \item \textit{Further studies of activities in multiple software development systems}.
    Our results show that forums are being actively used by contributors that are also active in the other systems. 
    This is indeed evidence that supports the claim that forums do work as a center of knowledge-sharing.
    Possible future directions could include tracking the actual threads to understand whether the discussions in forums cross-over into the review and bugs systems, and to what effect.
    
    \item \textit{Causal inference on promoting career}. 
    From Eclipse historical data of social roles, we can also collect contributor metrics of a leader. 
    Combined with the metrics of a committer, we can address the impact of the forum on social leadership, that is, the causal effect of forum participation on being promoted from committers to leaders.
\end{itemize}

\section{Threats to Validity}
\label{sec:threatstovalidity}
Several threats to the \textit{construct validity} emerge in our study.
To connect the dataset between five different sources, we used \textit{uids, email} addresses and \textit{names}. 
Although the inspection was conducted manually, it is possible that our approach might lead to ambiguity.
To mitigate this threat, we cross-reference between those five different systems to ensure that the authors are indeed not multiple aliases.
Related to the memberships classification, it is possible to have inaccuracy of threshold definition to identify the member statuses. 
Although the thresholds are not formally informed by the organization, we specified the thresholds based on the maximum number of posts found in our dataset for each member status.
Since the number of user misidentifications is small, we consider to use our defined thresholds in our study.
It is also possible that there are identical users that have different identity number and different status in the forums.
We also found in several threads, a message posted by a user had been duplicated with a different message identity number. 
Since we distinguished the users and the messages by their identity number, in our analyses, we recorded them differently even though they are same.
In relation with forums linkage, we found that not all links in the threads can be extracted since they are written in plain text by users.
However, the number of these issues were small.
Thus, we consider that the impact of the missing links is not significant.
The other threats to the construct validity is related to our manual labeling in RQ$_2$ and RQ$_3$. 
The label might be affected by annotator misunderstand or mislabeling. 
Despite annotators resolving disagreements through discussion, the labels might still be incorrect.
In addition, there are many potential factors need to be taken into account. To mitigate this threat, we limit the scope of our study into conversational contents of the discussions themselves, to gain the insights of how project-specific forums are utilized.

Threats to the \textit{external validity} appear in our data preparation. 
Even though we investigated a large scale of discussion forums in Eclipse, the findings could not be generalized to other organizations.

We diminish the threats to \textit{reliability} by preparing the online appendix of our collected threads, links, contributors, and the results of our manual annotation for RQ$_2$ and RQ$_3$ (see Section~\ref{ssec:appendix}).

\section{Related Work}
\label{sec:relatedwork}
Forum discussion has been a popular research topic in the software engineering community in recent years.
In this section, we introduce the key research related to our study.

\subsection{Stack Overflow.}
There are a number of studies on a web-based question-and-answer communication channel, Stack Overflow (SO).
The collaboration between users in different community forums~\citep{Zagalsky:2018:RCC:3211160.3211168} and the knowledge sources contained in the discussion forums~\citep{Ye:2017:SDK:3042021.3042050} have shaped the characteristics of knowledge sharing in the community.

\cite{Zagalsky:2018:RCC:3211160.3211168} reported that SO with R-tag has become one of the two questions and answers communication channels that have a relationship with the users' discussions in R software development community forum.
The collaboration between members and the independent individuals work have shaped the knowledge characteristics of the community.
\cite{Ye:2017:SDK:3042021.3042050} analyzed the reason behind the inclusion of web links in SO forum. 
It has been reported that the forum users share the URLs in SO for various purposes.
Referencing sources to provide the solutions is the most prevalent purposes for the users who posted the web links.

In our study, we analyzed the entire discussion for each topic (not per message) in the Eclipse forums.
Thus, we modified the label proposed by~\cite{Zagalsky:2018:RCC:3211160.3211168} to code the types of a thread (see Section~\ref{ssec:charofthreads}).
Compared to the work by~\cite{Ye:2017:SDK:3042021.3042050}, we found that enclosing link sources to report bug-related problems or find a software defect is common in the forums.

Other works have studied the factors that affect the movement of users to SO and the most common topics discussed.

\cite{Squire:2015:WMS:2819009.2819042} showed that the quality measurements, participation of
users, and the effectiveness of responding time in the SO forum have been the main factors that induced the developers' movement from self-supported forums and mailing lists.
\cite{Zou:2017:TCN:3044551.3068580} describes the insights of developers' requirements through SO.
The authors found that the usability and reliability are the most common topics discussed in the SO forum and also the most unresolved problems faced by the users.
% By visualizing the development activities evolution over time, most software developers interested in the functionality and reliability, while the usability still becoming the trends.

To complement with the previous work by~\cite{Squire:2015:WMS:2819009.2819042}, our study emphasizes that the users in Eclipse forums also actively participate and contribute over different systems in the Eclipse ecosystem.
Expanding the work by~\cite{Zou:2017:TCN:3044551.3068580}, we found that the most common topics discussed in Eclipse forums are Eclipse platforms and BIRT.

Several studies investigated on how to submit a successful question and increase the chance of receiving a successful answer in SO.

\cite{Calefato:2018:ATH:3163583.3163673} proposed guidelines for users on how to increase the chances of receiving a successful answer to their questions on Stack Overflow.
They found that the factors of potential-to-success questions are conciseness, completeness, and the exactness of the characters usage.
In terms of affect, a neutral emotional style may impact successful questions.
In line with that,~\cite{Wang:2018:UFF:3231288.3231332} investigated the factors related to the time to get an accepted answer in four Stack Exchange websites.
After controlling for other factors, the accepted answers had been affected strongly by the dimension of the answerers.
The more frequent a user is in answering the questions, the faster to receive an accepted answer.

In comparison with the work by~\cite{Calefato:2018:ATH:3163583.3163673} and~\cite{Wang:2018:UFF:3231288.3231332}, our study shows that it is important to ask a procedural question in a positive manner, since it might increase the chances of receiving responses.

\subsection{Twitter and news aggregators.}
Communication between programmers are not limited to only mailing lists and question-and-answer channels, but also frequently occur on social media (i.e. Twitter) and news aggregators.

The information hidden in the tweets posted by Twitter users have been attractive for researchers to explore in a number of studies. 

\cite{Mezouar:2018:TUB:3231288.3231333} investigated the tweets posted by users to relate to the bug reports after removing the noisy tweets. 
\cite{Guzman:7765515} analyzed the characteristics of the usage of Twitter, information in the messages, and classified automatically the potential messages about software applications. 

In our study, the findings support prior works by~\cite{Mezouar:2018:TUB:3231288.3231333} and~\cite{Guzman:7765515}. 
We show that bug report are the most dominant link target types referenced in the forums, while Eclipse Projects has become the most popular forums in the Eclipse community.
In addition, Eclipse forum is not only used for asking solutions, but also for sharing information or announcement.

\subsection{Contributors.}
Software development could not be separated from users' participation in a forum. 
Their contributions are not always related to writing code.
A number of studies on the contributors in communication channels, has shown that experienced contributors and newcomers play an important role in developing software.

\textbf{Senior contributors.}
Recent studies show that every individual has an opportunity to become a valuable contributor.
\cite{Zhou:2012:6227164} built a model to analyze the users' chances of becoming a senior contributor depend on her competence, passion, and first-time contribution opportunity. 
\cite{Zhou20156880395} found that the participation of new members in the issue tracking system environment might impact their status of becoming a long term contributor.
Expanding the findings from the two prior works, we found that becoming a senior contributor depends on the frequency of posts (i.e., Q\&A, general information, or announcement) in the Eclipse forums.

\textbf{Junior contributors.}
Despite most of the valuable information for improvement to software quality comes from the experienced members in a software project~\citep{Li:2015:MGS:2818754.2818839}, software developers should not underestimate the newcomers' contributions in a discussion forum.

\cite{STEINMACHER201567} identified that the lack of social interaction with the community, having unanswered questions or receiving delayed answers, and their technical experience backgrounds are some difficulties that new members faced when they make contributions to an open source software project.
\cite{Middleton:2018:CPW:3196398.3196429} also studied the contribution characteristics of new members in OSS projects. The authors identified that the participation forms of the new members, such as pull request and how they comment in the discussion influence the decision to join OSS teams. 
Our study extends the previous works that posting a message in positive manner might increase the chances of getting positive responses.
Thus, it will have an impact on increasing social interaction.

\section{Conclusion}
\label{sec:conclusion}
To understand the impact of Eclipse community forums related to the linkage of the ecosystem and the social leadership, we conducted (i) a statistical study of 832,058 threads to analyze the characteristics of threads, contributions of Eclipse users, and users' classification; and also (ii) a large-scale study on the membership-based participation, content, and a sentiment investigation on the social interaction.

Our empirical study has shown that forums are an essential platform for linking various resources in Eclipse ecosystem and a source of expert knowledge for other development systems. 
The results also reveal that forum members actively participate in posting and responding to threads without limitation of membership. 
The forums are dominated by question and answer threads, with the status of users affecting the contents. 
Sentiment of interaction among developers are inconsistent.
This work shows Eclipse forum is just one example of the utilization of project-specific forums.
We recommend the users join forums as they are the center of knowledge sharing.
For developers, preparing project-specific forums is important to share community-related information or announcement.
Our results also hint at researchers that there are many open issues for future work: understanding and supporting forum discussions, further studies of activities in multiple software development systems, and causal inference for assessment, to name a few.

%\begin{acknowledgements}
%\end{acknowledgements}

% \bibliography{reference}   % name your BibTeX data base

\end{document}